\documentclass[twocolumn]{aastex631}
\usepackage{graphicx}	
\usepackage{subfigure}

\shorttitle{Polarization towards NGC 7380}
\shortauthors{Singh et al.}

\graphicspath{{./}{figures/}}

\begin{document}

\title{Foreground Dust Properties towards the Cluster NGC 7380}

\correspondingauthor{Sadhana Singh}
\email{sadhana.ftp22@gmail.com}
\author{Sadhana Singh}
\affiliation{Astronomy \& Astrophysics Division, Physical Research Laboratory (PRL), Ahmedabad-380009, India} 
\affiliation{Aryabhatta Research Institute of Observational Sciences (ARIES), Manora Peak, Nainital 263001, India}

\author{Jeewan C. Pandey}
\affiliation{Aryabhatta Research Institute of Observational Sciences (ARIES), Manora Peak, Nainital 263001, India}

\author{Thiem Hoang}
\affiliation{Korea Astronomy and Space Science Institute, Daejeon 34055, Republic of Korea}
\affiliation{Department of Astronomy and Space Science, University of Science and Technology, 217 Gajeong-ro, Yuseong-gu, Daejeon, 34113, Republic of Korea}

\author{Neelam Panwar}
\affiliation{Aryabhatta Research Institute of Observational Sciences (ARIES), Manora Peak, Nainital 263001, India}

\author{Biman J. Medhi}
\affiliation{Gauhati University, Guwahati 781014, India}

\author{Vishal Joshi}
\affiliation{Astronomy \& Astrophysics Division, Physical Research Laboratory (PRL), Ahmedabad-380009, India}

\author{Shashikiran Ganesh}
\affiliation{Astronomy \& Astrophysics Division, Physical Research Laboratory (PRL), Ahmedabad-380009, India}

\begin{abstract}
Using starlight polarization, we present the properties of foreground dust towards cluster NGC 7380 embedded in H{\sc ii} region Sh 2-142. Observations of starlight polarization are carried out in four filters using an imaging polarimeter equipped with a 104-cm ARIES telescope. Polarization vectors of stars are aligned along the Galactic magnetic field. Towards the east and southeast regions, the dust structure appears much denser than in other regions (inferred from extinction contours and colour composite image) and is also reflected in polarization distribution. We find that the polarization degree and extinction tend to increase with distance and indication for the presence of a dust layer at a distance of around 1.2 $kpc$. We have identified eight potential candidates exhibiting intrinsic polarization by employing three distinct criteria to distinguish between stars of intrinsic polarization and interstellar polarized stars. For interstellar polarized stars, we find that the maximum polarization degree increases with the color excess and has a strong scatter, with the mean value of 1.71$\pm$0.57$\%$. The peak wavelength spans $0.40-0.88\mu$m with the mean value of 0.56$\pm$0.07 ~$\mu m$, suggesting similar grain sizes in the region as the average diffuse interstellar medium. The polarization efficiency is also found to decrease with visual extinction as $P_{max}/A_{V}\propto A_{V}^{-0.61}$. Our observational results are found to be consistent with the predictions by the radiative torque alignment theory.
\end{abstract}

.

\section{Introduction} \label{intro}
Interstellar dust is responsible for starlight extinction, and their re-emission in long wavelengths illuminates the Universe. Dust also plays a vital role in many astrophysical processes, including gas heating and cooling, star and planet formation, and surface chemistry (see \citealt{2011piim.book.....D} book). The study of the physical properties of interstellar dust (including size, shape, and composition) thus helps us better understand the astrophysical environments and constrain the role of dust in different physical processes. 

The leading methods to extract dust properties include the extinction and polarization of starlight. \citet{1949Sci...109..166H} and  \citet{1949Sci...109..165H, 1949ApJ...109..471H} first discovered the polarization of distant stars and suggested the polarization is caused by alignment of non-spherical dust grains with the ambient magnetic field. Such a discovery had opened a new avenue for tracing interstellar magnetic fields using starlight polarization \citep{1975ApJ...196..261S, 2000AJ....119..923H}. This technique is based on the fact that grains tend to align with their long axes perpendicular to the magnetic field, which results in the differential extinction and the net polarization of starlight with the polarization vectors (${\bf P}$) parallel to the magnetic field direction (see, e.g., \citealt{Hoang.2023}).

There have been several studies about the properties and alignment of interstellar dust using starlight polarization \citep[e.g.][]{2015ARA&A..53..501A}.
The size distribution of grains varies with the local conditions due to different growth \citep[e.g. accretion, coagulation;][]{2012MNRAS.422.1263H} and destruction processes. Observations of starlight polarization toward molecular clouds are valuable to constrain grain growth \citep{2020ApJ...905..157V}. 

The polarization of starlight relies on the systematic alignment of elongated dust grains with the ambient magnetic field. However, the physics of grain alignment and how grain alignment depends on grain composition and the local environments is not well understood (see \citealt{2015ARA&A..53..501A}; \citealt{2015psps.book...81L} for reviews). 
The classical theory of grain alignment proposed by \citet{1951ApJ...114..206D} (aka. Davis-Greenstein mechanism) is based on the paramagnetic relaxation of dust grains containing unpaired electrons, which are paramagnetic material. However, this mechanism is shown to be inefficient for large grains. \citet{1967ApJ...147..943J} improved the Davis-Greenstein mechanism by considering superparamagnetic grains containing embedded iron inclusions. \citet{1979ApJ...231..404P} suggested several surface processes that can spin up grains to suprathermal rotation and align grains. \citet{1952Natur.169..322G, 1952MNRAS.112..215G} proposed the mechanical alignment due to the interaction of gas flow with dust grains, but this mechanism requires supersonic flow to be efficient. The leading grain alignment theory is now based on radiative torques (RAT) arising from the interaction of left-handed and right-handed photons with irregular dust grains, first introduced by \citet{1976Ap&SS..43..291D} and numerically demonstrated in \citet{1996ApJ...470..551D, 1997ApJ...480..633D}. The analytical model of grain alignment based on RATs was introduced in \cite{2007MNRAS.378..910L}, which allowed us to conduct comprehensive studies on grain alignment for different environments \citep{2008MNRAS.388..117H, 2009ApJ...695.1457H, 2009ApJ...697.1316H}. Later studies found that grains with iron inclusions could be efficiently aligned by both the effects of magnetic relaxation and radiative torques, which is termed MRAT mechanism \citep{2016ApJ...831..159H, 2018ApJ...852..129H, 2021ApJ...908...12L, 2021ApJ...908..218H}. The latest observations by Planck for the interstellar medium (ISM) and by Atacama Large Millimeter/submillimeter Array (ALMA) for the dense regions around protostars favor the MRAT alignment mechanism \citep{2023MNRAS.520.3788G}.

Polarimetric observations are vital to probe grain properties and test grain alignment mechanisms. Spectropolarimetric observations show the strongly polarized silicate feature at 9.7$\mu m$, which indicates that silicate grains must be non-spherical and efficiently aligned \citep{1983A&A...117..289C, 1986ApJ...308..281M, 2012A&A...541A..52V, 2013AstL...39..421V}. However, the carbonaceous spectral feature at 3.4$\mu m$ is negligibly polarized \citep{2006ApJ...651..268C}, which implies that carbonaceous grains are weakly aligned or have spherical shapes \citep{Hoang.2023}. 
There are different models developed for the interpretation of interstellar polarization in which various sizes, shapes, and composition of dust grains are considered \citep[e.g. see the review by][]{2012JQSRT.113.2334V} and the efforts for the prediction of polarized dust emission have been made \citep{2009ApJ...696....1D, 2023ApJ...948...55H}.  

In our studies (e.g., \citealt{2022MNRAS.513.4899S} for NGC 2345), we aim to probe dust properties and grain alignment physics by observing the polarization of starlight from open star clusters. 
We note that open star clusters appear to be the best environment for studying dust properties and grain alignment using starlight polarization because of the similar properties of stars and the availability of photometric and spectroscopic data of stars in an open cluster. Thus, by adding this information to the polarimetric study, we can deeply analyse the dust grains' properties in the region. Besides, the study of the evolution of physical parameters of dust grains all over the region is possible because of the diverse spread area of the cluster. The additional observations of non-member stars in the same region are useful for investigating the dust distribution along the line of sight.
The location of open star clusters in the Galactic spiral arm is also a major advantage for studying dust's properties.

Our target chosen for this study is the open star cluster NGC 7380. The cluster [R.A. (J2000) = $22^{h}47^{m}21^{s}$, Dec. (J2000)= $58^\circ07{^\prime}54{^{\prime\prime}}$] is associated with the H{\sc ii} region Sh 2-142 \citep{1959ApJS....4..257S,1976A&AS...25...25D}. It is located at Galactic longitude ($l$) of 107\degr.128 and Galactic latitude ($b$) of -0\degr.884. 
The binary O-type star DH Cep is considered to trigger star formation in neighbouring molecular clouds. It is a primary exciting star for the H{\sc ii} region Sh 2-142. DH Cep is found to be close to the zero-age main sequence \citep{1994A&A...282...93S}. Sh 2-142 reveals the nebula's peculiar structure, which is made up of several bright arcuate clouds, rims, and $H\alpha$ knots interlaced with local dust lanes \citep{1994A&A...283..963C}. The cluster is well studied photometrically, and a deficit of faint stars in the centre was found \citep{1971A&A....13...30M, 2011AJ....142...71C}. An age of 4 Myr was found in \citet{2011AJ....142...71C}, while \cite{1971A&A....13...30M} found a 2 Myr age of the cluster and the distance has been found in the range of 1.4 to 3.6 kpc in previous studies \citep{1971A&A....13...30M, 1971A&AS....4..241B, 1994A&A...283..963C, 2002NewA....7..553T, 2005A&A...438.1163K, 2020A&A...633A..99C, 2023A&A...673A.114H}. A variable reddening with the average value [$E(B-V)$] of $\sim$ 0.50-0.65 mag has been obtained. \citet{2002NewA....7..553T} and \citet{1971A&AS....4..241B} have determined the diameter of the cluster to be 5${^\prime}$ to 9${^\prime}$, where as \citet{2005A&A...438.1163K} found the core and cluster radii as 6${^\prime}$ and 15${^\prime}$. The photometric study of NGC 7380 by \citet{2011AJ....142...71C} shows the elongation of the cluster in the north-south direction and extension of stellar distribution towards the west of the cluster. The average angular size is estimated as 4\arcmin\, radius for the cluster. In addition to this, two dust shells were also identified in the region at the radius of 6${^\prime}$.5 and 10${^\prime}$.4 by \citet{1971A&A....13...30M} using projected star density counts and a map of interstellar reddening. 
There has been one polarimetric study for this cluster by \citet{1976AJ.....81..970M} around five decades ago in which unfiltered polarimetry of 42 stars was presented. However, in the study, only 10 stars were presented for multi-colour polarimetry. 
The association of NGC 7380 with the H{\sc ii} region highlights the importance of a complete study of the cluster region by including more stars with multiband polarimetric data and detailed analysis to see dust grains' properties and their distribution. 

We organize our paper as follows: the observations and data reduction are given in Section \ref{obs}, the results and analysis are reported in Section \ref{res}, and the discussion and summary are described in Section \ref{diss} and \ref{summ}, respectively. 

\section{Observation and data reduction}\label{obs}
\subsection{Polarimetric Observations}
The polarimetric observation for the cluster NGC 7380 was carried out using the ARIES Imaging Polarimeter (AIMPOL) on 12 October and 16 December 2018, along with polarized (HD 19820, HD 25443) and unpolarized standard (HD 21447) stars. The instrument AIMPOL \citep{2004BASI...32..159R, 2023JAI....1240008P} is mounted as a backend instrument of the 104-cm telescope of ARIES. The telescope is a Ritchey-Chretien type with a focal ratio of f/13. The AIMPOL is a dual-beam polarimeter that consists of a rotatable half-wave plate (HWP) and a Wollaston prism. The observations are taken at four positions of HWP (i.e. 0$\degr$, 22$\degr$.5, 45$\degr$, and 67$\degr$.5 from the celestial north-south direction). To get information about the size distributions of dust grains, we have taken observations in $B$, $V$, $R$, and $I$ bands. Exposure times for each frame in  $B$, $V$, $R$, and $I$ were 300, 180, 90, and 90 seconds, respectively. To get a good signal-to-noise ratio, we have observed at least three frames for each position of HWP and in each band. The standard procedure for the data reduction (e.g. cleaning of images, alignment and image combination, aperture photometry) has been applied using the Image Reduction and Analysis Facility\footnote{http://iraf.net}. Further, detailed data reduction procedures are described in \citet{2004BASI...32..159R} and \citet{2020AJ....159...99S}. The polarization parameters are extracted at different apertures for each object, and then using the curve of growth, the polarization's best value is selected for each object. Stokes parameters of the unpolarized standard star were used for the instrumental and other systematic corrections. The instrumental polarization was found to be less than 0.3$\%$ in all passbands during the observations. The zero point polarization angle was corrected by comparing the results of the position angle values of polarized standard stars with the standard values. The degree of polarization ($P$) and position angle ($\theta$) for polarized standard stars in all four bands are found to be close to the standard values from literature \citep{1992AJ....104.1563S}.

All those faint sources with  $P/\sigma_{P} < 2$, located on edge in observed frames and/or overlapped other sources, were not considered in the observed region. 
Thus, after applying all these selection criteria in polarization data, 72 sources are included in this study. The astrometry of the observed sources was carried out using the online available tool\footnote{https://nova.astrometry.net/upload}.

\subsection{Complementary data from other catalogues}
To get the Gaia ID of observed sources, we have cross-matched our sources with the Gaia counterpart \citep{2021A&A...649A...1G, 2018A&A...616A...1G, 2016A&A...595A...1G} with position reference. All observed sources have a Gaia counterpart, with most sources being cross-matched within 1\arcsec\, and only ten being cross-matched between 1\arcsec\,  and 3\arcsec.2. 
We have verified that all the observed sources, cross-matched with Gaia, satisfy the five parameter conditions in \citet{2018A&A...616A...2L} for a good astrometric solution. 
The information about member stars in the observed region of the cluster is taken from \citet{2023A&A...673A.114H}, \citet{2021MNRAS.504..356D}, and \citet{2017MNRAS.470.3937S}. We found 27, 10, and 15 observed stars as members in \citet{2017MNRAS.470.3937S}, \citet{2021MNRAS.504..356D}, and \citet{2023A&A...673A.114H}, respectively, with eight common members in all. \citet{2021MNRAS.504..356D} and \citet{2023A&A...673A.114H} have used Gaia DR2 and Gaia DR3 data for the analysis; therefore, we have considered the observed star as a member, if that star is given as member in either of these two catalogs. Further, the selection of member stars for this work from the given members in \citet{2017MNRAS.470.3937S} is done using the Stokes plane and the distance information of the star. As a result, 18 stars are assigned as members of the cluster NGC 7380 out of 72 observed stars in the cluster region.

During this analysis, the extinction and distance for each star are the basic information required. 
The basic parameters of the cluster are extracted from the recent early data release (Gaia EDR3) in \citet{2022A&A...658A..91A} along with the other various photometric surveys like Two Micron All-Sky Survey ($2MASS$), Wide-field Infrared Survey ($AllWISE$), Panoramic Survey Telescope and Rapid Response System ($Pan-STARRS$), and $SkyMapper$, using the Bayesian algorithm, StarHorse. \citet{2022A&A...658A..91A} catalogue presents the re-calibrated values for parameters like distance, extinction, etc. for stars at different percentiles. We have taken the extinction and distance information of each observed star from \citet{2022A&A...658A..91A} by matching the Gaia identification Id of stars and took the median value (50$^{th}$ percentile) of extinction and distance for each star. 
A good number of stars are found with the value of extinction and distance in \citet{2022A&A...658A..91A} i.e. 60 out of 72 observed stars, and these stars are marked with `$^{a}$' symbol in Table \ref{tab:pthe}. The non-matching of the other 12 stars could be due to the conditions applied during the analysis in \citet{2022A&A...658A..91A} for the precise and correct values of parameters.  
We estimated the distance and extinction for the cluster NGC 7380 as $2.5\pm0.2$ $kpc$ and $1.5\pm0.3$ $mag$, respectively, by fitting the Gaussian curve to their distribution of members. The derived values of distance and extinction for the cluster are found to be close to as estimated in \citet{2011AJ....142...71C} and \citet{2023A&A...673A.114H}. These values are further used in the analysis where we need the average value of extinction and the distance for the cluster NGC 7380.
The R.A. and Dec. of the observed star along with the information of Gaia magnitude ($G$) \citep{2018A&A...616A...1G}, distance \citep{2022A&A...658A..91A}, extinction \citep{2022A&A...658A..91A}, membership \citep{2023A&A...673A.114H, 2021MNRAS.504..356D, 2017MNRAS.470.3937S}, and Stokes parameters\footnote{given in Section \ref{sec:qu}} are given in Table \ref{tab:all_data}.
\startlongtable
\begin{deluxetable*}{llccccccc}
\tabletypesize{\scriptsize}
\tablecaption{R.A. (J2000), Dec. (J2000), $G$ magnitude, distance, extinction, membership information, and Stokes parameters of observed stars. \label{tab:all_data}}
\tablehead{
\colhead{ID} & \colhead{R.A. (\degr)} & \colhead{Dec. (\degr)} & \colhead{$G$ ($mag$)} & \colhead{Distance ($kpc$)} & \colhead{$A_{V}$ ($mag$)} & \colhead{Member} & \colhead{$Q_{V}$ ($\%$)} & \colhead{$U_{V}$ ($\%$)}
}
\startdata
1 	&  342.040316  &  58.116110  &  12.734  &  -	   &	 -	 &	      &  -0.5328$\pm$0.3109  & 1.5299  $\pm$ 0.3101    \\
2 	&  341.998384  &  58.141382  &  11.744  &  2.688    &	 0.919 	 &	      &  -0.9030$\pm$0.0471  & 2.0281  $\pm$ 0.0494    \\
3 	&  341.960572  &  58.086864  &  10.491  &  2.555    &	 1.564 	 &	Yes   &  -1.2476$\pm$0.1271  & 2.7503  $\pm$ 0.1294    \\
4 	&  341.954676  &  58.077004  &  12.908  &  2.334    &	 0.949 	 &	Yes   &  -1.3225$\pm$0.0812  & 3.0561  $\pm$ 0.0802    \\
5 	&  341.947126  &  58.058256  &  11.828  &  1.274    &	 1.429 	 &	      &  -1.8394$\pm$0.2355  & 1.7516  $\pm$ 0.2351    \\
6 	&  341.946372  &  58.052000  &  12.137  &  -	   &	 -	 &	      &  -2.2009$\pm$0.0816  & 2.0239  $\pm$ 0.0819    \\
7 	&  341.939980  &  58.113638  &  11.610  &  -	   &	 -	 &	Yes   &  -1.2608$\pm$0.0416  & 2.0656  $\pm$ 0.0406    \\
8 	&  341.928207  &  58.142771  &  13.407  &  -	   &	 -	 &	      &  -0.5138$\pm$0.0618  & 0.6391  $\pm$ 0.0612    \\
9 	&  341.913188  &  58.159170  &  10.597  &  2.404    &	 1.487 	 &	Yes   &  -1.3504$\pm$0.1798  & 1.4481  $\pm$ 0.1799    \\
10	&  341.895869  &  58.126891  &  11.758  &  2.764    &	 1.506 	 &	Yes   &  -1.5912$\pm$0.0405  & 0.7285  $\pm$ 0.0423    \\
11	&  341.883733  &  58.161277  &  12.572  &  1.448    &	 1.246 	 &	      &  -1.1128$\pm$0.2168  & 1.6498  $\pm$ 0.2186    \\
12	&  341.878682  &  58.152364  &  13.799  &  2.643    &	 1.681 	 &	Yes   &  -1.5099$\pm$0.0483  & 0.8647  $\pm$ 0.0445    \\
13	&  341.869749  &  58.124267  &  13.550  &  2.630    &	 1.705 	 &	      &  -0.4242$\pm$0.0229  & 1.2605  $\pm$ 0.0204    \\
14	&  341.868163  &  58.103944  &  13.668  &  2.231    &	 1.275 	 &	Yes   &  -1.0543$\pm$0.2809  & 2.2005  $\pm$ 0.2802    \\
15	&  341.874591  &  58.099508  &  12.463  &  1.812    &	 1.326 	 &	      &  -1.0521$\pm$0.0894  & 1.8520  $\pm$ 0.0898    \\
16	&  341.883013  &  58.087019  &  11.594  &  -	   &	 -	 &	      &  -0.6020$\pm$0.0250  & 1.3520  $\pm$ 0.0211    \\
17	&  341.890490  &  58.068648  &  13.322  &  2.538    &	 1.827 	 &	Yes   &  -1.6881$\pm$0.0486  & 1.4571  $\pm$ 0.0481    \\
18	&  341.904324  &  58.063856  &  13.405  &  0.470    &	 0.861 	 &	      &  -1.1708$\pm$0.2006  & 0.4921  $\pm$ 0.2033    \\
19	&  341.836097  &  58.130844  &  12.526  &  2.538    &	 1.278 	 &	Yes   &  -1.5237$\pm$0.1716  & 1.4011  $\pm$ 0.1719    \\
20	&  341.837345  &  58.097383  &  13.822  &  2.446    &	 1.020	 &	Yes   &  -1.6104$\pm$0.0297  & 1.3322  $\pm$ 0.0295    \\
21	&  341.837738  &  58.088919  &  13.245  &  2.649    &	 1.484 	 &	Yes   &  -1.4252$\pm$0.0683  & 1.5829  $\pm$ 0.0686    \\
22	&  341.853906  &  58.075045  &  14.310  &  2.591    &	 1.454 	 &	      &  -3.0021$\pm$0.0498  & 1.8759  $\pm$ 0.0496    \\
23	&  341.956225  &  58.147196  &  10.078  &  1.468    &	 1.370 	 &	      &   -    & -       \\
24	&  341.928770  &  58.110909  &  19.923  & -	   &	 -	 &	      &  -0.4917$\pm$0.0802  & -0.4097 $\pm$ 0.0803    \\
25	&  341.901222  &  58.169802  &  13.177  &  2.545    &	 1.816 	 &	Yes   &  -1.1931$\pm$0.0101  & 0.2018  $\pm$ 0.0126    \\
26	&  341.862178  &  58.265235  &  14.421  &  2.696    &	 2.307 	 &	      &  -1.5394$\pm$0.7804  & 0.8323  $\pm$ 0.7815    \\
27	&  341.837084  &  58.221637  &  12.553  &  1.429    &	 1.563 	 &	      &  -1.7046$\pm$0.1198  & 1.2938  $\pm$ 0.1197    \\
28	&  341.839683  &  58.203090  &  12.580  &  0.656    &	 0.836 	 &	      &  -0.8168$\pm$0.1520  & 1.0999  $\pm$ 0.1511    \\
29	&  341.859950  &  58.218626  &  11.460  &  2.509    &	 1.707 	 &	Yes   &  -1.7080$\pm$0.1300  & 0.0835  $\pm$ 0.1373    \\
30	&  341.901654  &  58.187579  &  10.215  &  0.305    &	 0.231 	 &	      &   -     & -      \\
31	&  341.874470  &  58.187318  &  14.298  &  1.541    &	 1.624 	 &	      &  -0.5249$\pm$0.0480  & 1.4265  $\pm$ 0.0497    \\
32	&  341.887660  &  58.216920  &  15.319  &  3.812    &	 2.761 	 &	      &   1.2744$\pm$0.4691  & -1.3957 $\pm$ 0.4693    \\
33	&  341.903309  &  58.265643  &  11.955  &  4.474    &	 2.181 	 &	      &  -0.6975$\pm$0.0920  & 0.9887  $\pm$ 0.0910    \\
34	&  341.917302  &  58.229111  &  12.802  &  2.474    &	 1.491 	 &	      &  -0.6600$\pm$0.2191  & 0.4866  $\pm$ 0.2184    \\
35	&  341.923090  &  58.220603  &  13.282  &  0.536    &	 0.927 	 &	      &   -     & -       \\
36	&  341.915761  &  58.204857  &  14.378  &  1.987    &	 2.146 	 &	      &  -1.5304$\pm$0.2400  & 0.1717  $\pm$ 0.2366    \\
37	&  341.953932  &  58.219837  &  13.668  &  0.391    &	 0.698 	 &	      &  -0.6142$\pm$0.0105  & 1.3795  $\pm$ 0.0101    \\
38	&  341.953584  &  58.213372  &  14.120  &  2.987    &	 2.836 	 &	Yes   &  -1.7538$\pm$0.0900  & 0.1473  $\pm$ 0.0860    \\
39	&  341.953382  &  58.195011  &  14.441  &  2.716    &	 2.011 	 &	Yes   &  -2.0368$\pm$0.0999  & 0.3080  $\pm$ 0.0936    \\
40	&  341.970568  &  58.254639  &  11.857  &  2.961    &	 2.116 	 &	      &  -1.1140$\pm$0.1098  & 1.1063  $\pm$ 0.1098    \\
41	&  341.956041  &  58.284074  &  13.189  &  0.542    &	 0.480 	 &	      &  -0.5038$\pm$0.0585  & 0.7698  $\pm$ 0.0594    \\
42	&  342.035295  &  58.276696  &  13.822  &  -	   &	 -	 &	      &  -0.7715$\pm$0.3196  & 1.5819  $\pm$ 0.3199    \\
43	&  342.030679  &  58.260169  &  13.229  &  0.718    &	 1.128 	 &	      &  -0.4057$\pm$0.0932  & 0.8701  $\pm$ 0.0907    \\
44	&  342.035291  &  58.236029  &  12.285  &  1.619    &	 1.359 	 &	      &  -1.2586$\pm$0.1985  & 1.4894  $\pm$ 0.1989    \\
45	&  342.003013  &  58.206309  &  12.651  &  1.132    &	 1.446 	 &	      &  -0.5567$\pm$0.0218  & 1.1415  $\pm$ 0.0204    \\
46	&  341.994203  &  58.195861  &  13.095  &  1.028    &	 1.823 	 &	      &  -0.9829$\pm$0.0900  & 0.3080  $\pm$ 0.0899    \\
47	&  342.007743  &  58.201032  &  14.849  &  0.904    &	 1.543 	 &	      &  -0.7376$\pm$0.1593  & 1.1187  $\pm$ 0.1597    \\
48	&  341.944937  &  58.250959  &  11.318  &  2.074    &	 1.850 	 &	      &  -1.4956$\pm$0.1199  & 0.8290  $\pm$ 0.1195    \\
49	&  341.938515  &  58.231768  &  13.043  &  2.057    &	 1.269 	 &	      &  -0.8130$\pm$0.0105  & 0.3860  $\pm$ 0.0121    \\
50	&  341.624733  &  58.244880  &  12.242  &  0.299    &	 0.414 	 &	      &  -0.4460$\pm$0.0105  & -0.6276 $\pm$ 0.0103    \\
51	&  341.582157  &  58.202995  &  12.232  &  1.178    &	 1.244 	 &	      &  -0.6635$\pm$0.2315  & 0.4646  $\pm$ 0.2331    \\
52	&  341.649136  &  58.223632  &  13.604  &  -	   &	 -	 &	      &  -1.6527$\pm$0.6039  & 0.6610  $\pm$ 0.6238    \\
53	&  341.656935  &  58.196546  &  13.217  &  1.698    &	 0.985 	 &	      &  -1.5592$\pm$0.1200  & -0.0490 $\pm$ 0.1198    \\
54	&  341.660418  &  58.176325  &  13.264  &  1.766    &	 1.035 	 &	      &  -1.2903$\pm$0.0608  & 1.2116  $\pm$ 0.0610    \\
55	&  341.596170  &  58.149175  &  11.806  &  0.989    &	 1.215 	 &	      &  -0.3876$\pm$0.0300  & 0.6769  $\pm$ 0.0300    \\
56	&  341.598972  &  58.128154  &  14.296  &  4.739    &	 2.204 	 &	      &   -     & -       \\
57	&  341.568572  &  58.144122  &  13.008  &  1.068    &	 1.259 	 &	      &  -0.4599$\pm$0.0681  & 0.9104  $\pm$ 0.0695    \\
58	&  341.553969  &  58.155683  &  13.730  &  -	   &	 -	 &	      &  -0.7023$\pm$0.8312  & 2.3263  $\pm$ 0.8301    \\
59	&  341.642615  &  58.145746  &  14.095  &  2.376    &	 1.461	 &	Yes   &  -1.5765$\pm$0.4524  & 0.4520  $\pm$ 0.4786    \\
60	&  341.727041  &  58.224122  &  14.378  &  2.557    &	 1.897 	 &	Yes   &   -     & -       \\
61	&  341.727532  &  58.195256  &  12.954  &  0.565    &	 0.564 	 &	      &  -0.6045$\pm$0.2079  & 1.2071  $\pm$ 0.2095    \\
62	&  341.713353  &  58.170818  &  11.918  &  1.021    &	 0.943 	 &	      &  -0.8998$\pm$0.0994  & 1.0064  $\pm$ 0.0995    \\
63	&  341.677358  &  58.134852  &  12.677  &  2.609    &	 1.165 	 &	      &  -1.0827$\pm$0.1098  & 1.1369  $\pm$ 0.1098    \\
64	&  341.689767  &  58.128746  &  12.470  &  -	   &	-	 &	      &  -1.2142$\pm$0.1400  & -0.1191 $\pm$ 0.1405    \\
65	&  341.711188  &  58.140533  &  12.894  &  -	   &	-	 &	      &   -    & -      \\
66	&  341.720141  &  58.143702  &  11.747  &  2.154    &	 1.134 	 &	Yes   &  -1.6755$\pm$0.2097  & 1.4565  $\pm$ 0.2096    \\
67	&  341.736770  &  58.171898  &  14.442  &  1.861    &	 1.103 	 &	      &  -1.0986$\pm$0.4697  & 0.9154  $\pm$ 0.4695    \\
68	&  341.710826  &  58.234877  &  12.931  &  -	   &	 -	 &	      &  -0.7077$\pm$0.0100  & -0.0569 $\pm$ 0.0148    \\
69	&  341.744984  &  58.189533  &  13.363  &  0.507    &	 0.543 	 &	      &  -0.5062$\pm$0.2379  & 1.3907  $\pm$ 0.2397    \\
70	&  341.779124  &  58.211409  &  14.081  &  2.435    &	 2.170 	 &	      &   -     & -      \\
71	&  341.694154  &  58.144663  &  14.499  &  4.629    &	 2.337 	 &	      &  -1.1195$\pm$0.2538  & 4.1206  $\pm$ 0.2596    \\
72	&  341.736334  &  58.191529  &  14.540  &  0.833    &	 1.246 	 &  	      &   -     & -      \\ 
\enddata
\end{deluxetable*}

\section{Results and Analysis}\label{res}
Table \ref{tab:pthe} describes the polarization and position angle in all four bands for observed sources in the region of the cluster NGC 7380. The first column is the Id, whereas the second and third columns are the identification Id from \textit{Gaia} and offset between \textit{Gaia} positions and observed positions. The consecutive columns are the polarization value and angle in $B$, $V$, $R$, and $I$ bands. The cluster member stars are marked with the asterisk symbol with their Id.   

\startlongtable
\begin{deluxetable*}{llccccccccc}
\tabletypesize{\scriptsize}
\tablecaption{Degree of polarization and polarization angle for observed sources in $B$, $V$, $R$, and $I$ bands in the region of the cluster NGC 7380.\label{tab:pthe}}
\tablehead{
\colhead{ID} & \colhead{\textit{Gaia} ID} & \colhead{Offset(${^{\prime\prime}}$)} & \colhead{$P_{B}\left(\%\right)$} & \colhead{$\theta_{B}\left(^\circ\right)$} & \colhead{$P_{V}\left(\%\right)$} & \colhead{$\theta_{V}\left(^\circ\right)$} & \colhead{$P_{R}\left(\%\right)$} & \colhead{$\theta_{R}\left(^\circ\right)$} & \colhead{$P_{I}\left(\%\right)$} & \colhead{$\theta_{I}\left(^\circ\right)$}
}
\startdata
1 	&   2007420576900570496 & 0.76   & 2.06$\pm$0.17 &  57.5$\pm$  2.4   &  1.62$\pm$0.31 &   54.6$\pm$  5.5 &  1.94$\pm$0.06 & 60.0$\pm$  0.8  &   1.88$\pm$0.14 &  55.1$\pm$  2.1    \\	
2$^{a}$ 	&   2007421367174517120 & 0.69   & 1.55$\pm$0.13 &  56.7$\pm$  2.3   &  2.22$\pm$0.05 &   57.0$\pm$  0.6 &  2.04$\pm$0.10 & 64.3$\pm$  1.4  &   1.81$\pm$0.01 &  59.4$\pm$  0.1    \\
3*$^{a}$ 	&   2007418961992863744 & 0.55   & 2.77$\pm$0.08 &  56.2$\pm$  0.8   &  3.02$\pm$0.13 &   57.2$\pm$  1.2 &  2.85$\pm$0.06 & 59.4$\pm$  0.6  &   2.83$\pm$0.09 &  55.2$\pm$  0.9    \\
4*$^{a}$ 	&   2007418927633130752 & 0.37   & 2.83$\pm$0.18 &  57.8$\pm$  1.9   &  3.33$\pm$0.08 &   56.7$\pm$  0.7 &  3.14$\pm$0.02 & 59.5$\pm$  0.1  &   3.01$\pm$0.20 &  56.8$\pm$  1.9    \\
5$^{a}$ 	&   2007418721474713728 & 0.19   & 2.45$\pm$0.25 &  68.8$\pm$  2.9   &  2.54$\pm$0.24 &   68.2$\pm$  2.6 &  2.44$\pm$0.09 & 72.2$\pm$  1.1  &   2.27$\pm$0.04 &  68.3$\pm$  0.5    \\
6 	&   2007418652755241856 & 0.14   & 3.08$\pm$0.04 &  70.1$\pm$  0.3   &  2.99$\pm$0.08 &   68.7$\pm$  0.8 &  2.88$\pm$0.01 & 74.8$\pm$  0.1  &   2.45$\pm$0.01 &  74.6$\pm$  0.2    \\
7* 	&   2007419820986293504 & 0.44   & 2.51$\pm$0.06 &  63.8$\pm$  0.7   &  2.42$\pm$0.04 &   60.7$\pm$  0.5 &  2.26$\pm$0.03 & 69.0$\pm$  0.3  &   2.05$\pm$0.06 &  66.0$\pm$  0.8    \\
8 	&   2007420130223910400 & 0.54   & 0.90$\pm$0.31 &  50.3$\pm$  9.9   &  0.82$\pm$0.06 &   64.4$\pm$  2.2 &  1.05$\pm$0.11 & 66.0$\pm$  2.9  &   0.29$\pm$0.03 & 111.9$\pm$  3.1    \\
9*$^{a}$ 	&   2007423050801646720 & 0.54   & 1.80$\pm$0.06 &  70.4$\pm$  1.0   &  1.98$\pm$0.18 &   66.5$\pm$  2.6 &  1.94$\pm$0.02 & 76.6$\pm$  0.3  &   1.62$\pm$0.02 &  63.3$\pm$  0.4    \\
10*$^{a}$	&   2007419992784958464 & 0.39   & 1.86$\pm$0.15 &  75.2$\pm$  2.3   &  1.75$\pm$0.04 &   77.7$\pm$  0.7 &  1.92$\pm$0.11 & 82.6$\pm$  1.7  &   1.57$\pm$0.14 &  79.7$\pm$  2.5    \\
11$^{a}$	&   2007423325679536896 & 0.55   & 1.78$\pm$0.05 &  68.2$\pm$  0.8   &  1.99$\pm$0.22 &   62.0$\pm$  3.1 &  1.78$\pm$0.03 & 66.2$\pm$  0.4  &   1.46$\pm$0.01 &  65.8$\pm$  0.1    \\
12*$^{a}$	&   2007423016441899904 & 0.51   & 2.12$\pm$0.39 &  79.7$\pm$  5.2   &  1.74$\pm$0.05 &   75.1$\pm$  0.7 &  1.82$\pm$0.08 & 75.1$\pm$  1.3  &   1.70$\pm$0.16 &  77.4$\pm$  2.8    \\
13$^{a}$	&   2007419988475076736 & 0.22   & 1.17$\pm$0.16 &  46.5$\pm$  3.8   &  1.33$\pm$0.02 &   54.3$\pm$  0.5 &  1.23$\pm$0.08 & 68.8$\pm$  1.8  &   1.21$\pm$0.23 &  61.9$\pm$  5.5    \\
14*$^{a}$	&   2007419889705753472 & 0.32   & 1.41$\pm$0.06 &  63.4$\pm$  1.2   &  2.44$\pm$0.28 &   57.8$\pm$  3.3 &  2.23$\pm$0.07 & 66.5$\pm$  0.8  &   1.78$\pm$0.01 &  57.8$\pm$  0.1    \\
15$^{a}$	&   2007419885395865472 & 0.27   & 1.90$\pm$0.14 &  68.6$\pm$  2.0   &  2.13$\pm$0.09 &   59.8$\pm$  1.2 &  2.16$\pm$0.02 & 63.3$\pm$  0.3  &   2.08$\pm$0.01 &  59.2$\pm$  0.2    \\
16	&   2007419679237436800 & 0.30   & 1.50$\pm$0.07 &  57.1$\pm$  1.3   &  1.48$\pm$0.02 &   57.0$\pm$  0.5 &  1.63$\pm$0.05 & 65.8$\pm$  0.9  &   1.43$\pm$0.08 &  60.8$\pm$  1.7    \\
17*$^{a}$	&   2007418858913640960 & 0.26   & 2.46$\pm$0.43 &  67.7$\pm$  5.0   &  2.23$\pm$0.05 &   69.6$\pm$  0.6 &  2.61$\pm$0.02 & 69.1$\pm$  0.3  &   2.46$\pm$0.10 &  62.4$\pm$  1.2    \\
18$^{a}$	&   2007418824553909376 & 0.33   & 1.22$\pm$0.12 &  54.7$\pm$  2.9   &  1.27$\pm$0.20 &   78.6$\pm$  4.6 &  0.94$\pm$0.14 & 64.9$\pm$  4.3  &   0.54$\pm$0.13 &  57.2$\pm$  6.8    \\
19*$^{a}$	&   2007422982082164352 & 0.36   &        -      &        -          &  2.07$\pm$0.17 &   68.7$\pm$  2.4 &  2.03$\pm$0.05 & 68.2$\pm$  0.7  &   1.89$\pm$0.05 &  62.6$\pm$  0.8    \\
20*$^{a}$	&   2007419580468103296 & 0.26   & 2.24$\pm$0.21 &  75.3$\pm$  2.7   &  2.09$\pm$0.03 &   70.2$\pm$  0.4 &  2.13$\pm$0.03 & 68.5$\pm$  0.4  &   1.23$\pm$0.36 &  67.9$\pm$  8.3    \\
21*$^{a}$	&   2007419511748633856 & 0.27   & 1.80$\pm$0.17 &  62.9$\pm$  2.7   &  2.13$\pm$0.07 &   66.0$\pm$  0.9 &  2.04$\pm$0.08 & 68.2$\pm$  1.1  &   1.61$\pm$0.03 &  63.2$\pm$  0.5    \\
22$^{a}$	&   2007419305590219904 & 0.23   & 3.76$\pm$0.13 &  77.0$\pm$  1.0   &  3.54$\pm$0.05 &   74.0$\pm$  0.4 &  2.85$\pm$0.22 & 70.0$\pm$  2.3  &   2.06$\pm$0.11 &  61.0$\pm$  1.5    \\
23$^{a}$	&   2007421534663348608 & 0.72   & 1.82$\pm$0.08 &  43.9$\pm$  1.2   &  	  -         &   	-      &          -     &      -          &   1.56$\pm$0.08 &  43.1$\pm$  1.4    \\
24	&   2007419816682462464 & 1.61   & 1.37$\pm$0.28 & 165.8$\pm$  5.7   &  0.64$\pm$0.08 &  109.9$\pm$  3.6 &  1.06$\pm$0.15 &148.0$\pm$  4.1  &   0.77$\pm$0.25 & 106.9$\pm$  9.4    \\
25*$^{a}$	&   2007423325679749248 & 0.55   & 1.05$\pm$0.01 & 103.8$\pm$  0.1   &  1.21$\pm$0.01 &   85.2$\pm$  0.3 &  1.44$\pm$0.01 & 91.7$\pm$  0.1  &   1.44$\pm$0.03 & 101.6$\pm$  0.6    \\	
26$^{a}$	&   2007436348020244992 & 0.36   & 3.78$\pm$0.39 &  97.5$\pm$  3.0   &  1.75$\pm$0.78 &   75.8$\pm$ 12.8 &  1.44$\pm$0.29 & 68.8$\pm$  5.8  &   1.36$\pm$0.03 &  61.7$\pm$  0.6    \\
27$^{a}$	&   2007424386521626624 & 0.02   & 1.85$\pm$0.50 &  77.3$\pm$  7.7   &  2.14$\pm$0.12 &   71.4$\pm$  1.6 &  2.01$\pm$0.06 & 69.7$\pm$  0.9  &   1.68$\pm$0.12 &  65.8$\pm$  2.1    \\
28$^{a}$	&   2007424322111878272 & 0.49   & 1.47$\pm$0.15 &  72.4$\pm$  3.0   &  1.37$\pm$0.15 &   63.3$\pm$  3.2 &  1.23$\pm$0.02 & 66.8$\pm$  0.4  &   1.12$\pm$0.04 &  60.8$\pm$  1.0    \\
29*$^{a}$	&   2007424420881365504 & 0.41   & 1.73$\pm$0.06 &  85.5$\pm$  0.9   &  1.71$\pm$0.13 &   88.6$\pm$  2.3 &  1.79$\pm$0.10 & 89.6$\pm$  1.5  &   1.42$\pm$0.08 &  82.9$\pm$  1.7    \\
30$^{a}$	&   2007423428758730880 & 0.46   & 0.33$\pm$0.12 &  51.3$\pm$ 10.2   &  	       -    &        -         &        -       &     -           &   0.34$\pm$0.10 &  67.9$\pm$  8.7    \\
31$^{a}$	&   2007424150313468800 & 0.65   & 3.16$\pm$0.27 &  55.2$\pm$  2.4   &  1.52$\pm$0.05 &   55.1$\pm$  0.9 &  1.12$\pm$0.21 & 73.6$\pm$  5.4  &   1.61$\pm$0.19 &  67.4$\pm$  3.3    \\
32$^{a}$	&   2007424253392409984 & 0.02   &        -      &      -            &  1.89$\pm$0.47 &  156.2$\pm$  7.1 &  1.34$\pm$0.45 & 84.2$\pm$  9.6  &   1.97$\pm$0.15 & 119.8$\pm$  2.2    \\
33$^{a}$	&   2007437752459849856 & 0.33   & 1.15$\pm$0.12 &  54.4$\pm$  3.0   &  1.21$\pm$0.09 &   62.6$\pm$  2.2 &  1.25$\pm$0.07 & 69.6$\pm$  1.7  &   1.08$\pm$0.17 &  61.2$\pm$  4.6    \\
34$^{a}$	&   2007425730861160320 & 0.34   & 1.25$\pm$0.35 &  78.6$\pm$  8.0   &  0.82$\pm$0.22 &   71.8$\pm$  7.6 &  1.16$\pm$0.07 & 76.0$\pm$  1.6  &   0.86$\pm$0.16 &  58.6$\pm$  5.2    \\
35$^{a}$	&   2007425662141700736 & 0.28   & 1.07$\pm$0.24 &  53.9$\pm$  6.5   &      -         &       -          &  1.04$\pm$0.01 & 73.8$\pm$  0.1  &   0.97$\pm$0.03 &  61.5$\pm$  1.0    \\
36$^{a}$	&   2007423428758720000 & 0.27   & 2.87$\pm$0.19 & 103.3$\pm$  1.9   &  1.54$\pm$0.24 &   86.8$\pm$  4.4 &  2.05$\pm$0.01 & 94.8$\pm$  0.2  &   1.04$\pm$0.11 &  58.2$\pm$  3.0    \\
37$^{a}$	&   2007424940587209088 & 0.63   &         -     &       -           &  1.51$\pm$0.01 &   57.0$\pm$  0.2 &  1.14$\pm$0.03 & 63.4$\pm$  0.7  &   0.55$\pm$0.03 &  78.2$\pm$  1.5    \\
38*$^{a}$	&   2007424940587215232 & 0.27   & 2.95$\pm$0.07 &  94.8$\pm$  0.6   &  1.76$\pm$0.09 &   87.6$\pm$  1.4 &  1.43$\pm$0.21 & 91.6$\pm$  4.2  &   2.02$\pm$0.23 &  82.6$\pm$  3.2    \\
39*$^{a}$	&   2007424837508018560 & 0.58   & 1.57$\pm$0.21 &  94.7$\pm$  3.8   &  2.06$\pm$0.10 &   85.7$\pm$  1.3 &  2.26$\pm$0.01 & 88.9$\pm$  0.1  &   2.06$\pm$0.04 &  83.3$\pm$  0.6    \\
40$^{a}$	&   2007425863990374912 & 0.08   & 2.18$\pm$0.43 &  76.2$\pm$  5.7   &  1.57$\pm$0.11 &   67.6$\pm$  2.0 &  1.58$\pm$0.03 & 76.1$\pm$  0.6  &   1.37$\pm$0.14 &  70.5$\pm$  2.9    \\
41$^{a}$	&   2007437894208496896 & 0.55   & 0.80$\pm$0.09 &  44.6$\pm$  3.2   &  0.92$\pm$0.06 &   61.6$\pm$  1.8 &  1.24$\pm$0.12 & 64.3$\pm$  2.8  &   0.96$\pm$0.01 &  59.8$\pm$  0.1    \\
42	&   2007426276309283712 & 0.82   & 1.53$\pm$0.15 & 105.4$\pm$  2.9   &  1.76$\pm$0.32 &   58.0$\pm$  5.2 &  1.86$\pm$0.12 & 61.1$\pm$  1.9  &   2.07$\pm$0.28 &  60.8$\pm$  3.9    \\
43$^{a}$	&   2007426211897522816 & 0.27   & 1.14$\pm$0.24 &  67.1$\pm$  6.0   &  0.96$\pm$0.09 &   57.5$\pm$  2.8 &  1.61$\pm$0.09 & 62.2$\pm$  1.6  &   0.92$\pm$0.13 &  48.7$\pm$  4.2    \\
44$^{a}$	&   2007425004996916608 & 0.53   & 2.03$\pm$0.36 &  76.5$\pm$  5.1   &  1.95$\pm$0.20 &   65.1$\pm$  2.9 &  1.82$\pm$0.07 & 67.9$\pm$  1.2  &   1.65$\pm$0.01 &  62.7$\pm$  0.1    \\
45$^{a}$	&   2007424768788555008 & 0.77   & 1.45$\pm$0.16 &  48.7$\pm$  3.1   &  1.27$\pm$0.02 &   58.0$\pm$  0.5 &  1.24$\pm$0.05 & 62.3$\pm$  1.2  &   1.03$\pm$0.14 &  49.6$\pm$  3.8    \\
46$^{a}$	&   2007424700069085056 & 0.86   & 0.76$\pm$0.22 &  82.1$\pm$  8.3   &  1.03$\pm$0.09 &   81.3$\pm$  2.5 &  0.94$\pm$0.02 & 90.5$\pm$  0.6  &   0.86$\pm$0.09 &  74.3$\pm$  3.0    \\
47$^{a}$	&   2007424700069085824 & 1.04   &       -       &      -            &  1.34$\pm$0.16 &   61.7$\pm$  3.4 &          -     &         -       &	-         &	 -             \\  
48$^{a}$	&   2007425971379315840 & 0.32   & 1.44$\pm$0.02 &  77.7$\pm$  0.3   &  1.71$\pm$0.12 &   75.5$\pm$  2.0 &  1.46$\pm$0.11 & 81.0$\pm$  2.2  &   1.13$\pm$0.08 &  72.4$\pm$  2.0    \\
49$^{a}$	&   2007425730861170432 & 2.39   & 1.32$\pm$0.20 &  90.1$\pm$  4.3   &  0.90$\pm$0.01 &   77.3$\pm$  0.4 &  1.28$\pm$0.09 & 80.7$\pm$  2.0  &   1.06$\pm$0.19 &  75.9$\pm$  5.3    \\
50$^{a}$	&   2007431121030345728 & 0.51   & 0.71$\pm$0.18 & 173.8$\pm$  7.7   &  0.77$\pm$0.01 &  117.3$\pm$  0.4 &  0.80$\pm$0.07 & 78.1$\pm$  2.5  &   0.55$\pm$0.16 &  64.5$\pm$  8.3    \\
51$^{a}$	&   2007430914871918208 & 1.27   & 0.96$\pm$0.34 &  82.9$\pm$ 10.3   &  0.81$\pm$0.23 &   72.5$\pm$  8.3 &  0.57$\pm$0.14 & 72.3$\pm$  7.2  &   0.54$\pm$0.01 &  63.6$\pm$  0.8    \\
52	&   2007430850462052096 & 0.36   & 2.66$\pm$0.30 &  62.9$\pm$  3.3   &  1.78$\pm$0.60 &   79.1$\pm$ 10.1 &  1.88$\pm$0.13 & 76.4$\pm$  2.0  &   1.37$\pm$0.31 &  65.8$\pm$  6.4    \\
53$^{a}$	&   2007430021531203712 & 0.61   & 1.43$\pm$0.42 &  99.1$\pm$  8.5   &  1.56$\pm$0.12 &   90.9$\pm$  2.2 &  1.73$\pm$0.07 & 81.3$\pm$  1.2  &   1.16$\pm$0.04 &  79.6$\pm$  1.1    \\
54$^{a}$	&   2007429750950482048 & 0.68   & 1.46$\pm$0.53 &  53.9$\pm$ 10.5   &  1.77$\pm$0.06 &   68.4$\pm$  1.0 &  1.67$\pm$0.03 & 68.6$\pm$  0.4  &   1.46$\pm$0.11 &  78.4$\pm$  2.2    \\
55$^{a}$	&   2007429613511527296 & 0.57   & 0.75$\pm$0.11 &  59.7$\pm$  4.0   &  0.78$\pm$0.03 &   59.9$\pm$  1.1 &  0.68$\pm$0.05 & 58.6$\pm$  2.2  &   0.64$\pm$0.14 &  62.9$\pm$  6.2    \\
56$^{a}$	&   2007429510432338816 & 0.32   &       -       &      -  	   &  	     -	    &         -        &  1.04$\pm$0.49 & 76.5$\pm$ 13.7  &   1.61$\pm$0.36 &  29.4$\pm$  6.4    \\
57$^{a}$	&   2007429785310212352 & 0.81   &        -      &       - 	   &  1.02$\pm$0.07 &   58.4$\pm$  1.9 &  0.48$\pm$0.01 &179.8$\pm$  0.5  &   0.48$\pm$0.24 &  63.0$\pm$ 14.4    \\
58	&   2007429476072548224 & 3.13   & 1.75$\pm$0.10 &  49.3$\pm$  1.7   &  2.43$\pm$0.83 &   53.4$\pm$  9.8 &        -       &       -         &   1.58$\pm$0.50 &  41.1$\pm$  9.1    \\
59*$^{a}$	&   2007429647871293952 & 0.72   & 1.28$\pm$0.39 &  77.3$\pm$  8.9   &  1.64$\pm$0.45 &   82.0$\pm$  8.4 &  1.15$\pm$0.14 & 76.5$\pm$  3.6  &   1.53$\pm$0.03 &  68.3$\pm$  0.7    \\
60*$^{a}$	&   2007436038782582912 & 0.86   &       -       &      -  	   &  	      -     &          -       &  1.21$\pm$0.17 & 90.8$\pm$  4.0  &   1.07$\pm$0.02 &  95.2$\pm$  0.6    \\
61$^{a}$	&   2007435660825490944 & 0.70   & 1.20$\pm$0.33 &  54.0$\pm$  8.0   &  1.35$\pm$0.21 &   58.3$\pm$  4.4 &  1.03$\pm$0.07 & 65.1$\pm$  2.0  &   0.96$\pm$0.10 &  66.5$\pm$  3.1    \\
62$^{a}$	&   2007423836765817088 & 0.83   & 1.75$\pm$0.02 &  76.1$\pm$  0.3   &  1.35$\pm$0.10 &   65.9$\pm$  2.1 &  1.73$\pm$0.01 & 75.5$\pm$  0.1  &   1.42$\pm$0.07 &  65.0$\pm$  1.4    \\
63$^{a}$	&   2007417862481063424 & 0.36   & 1.15$\pm$0.17 &  70.4$\pm$  4.3   &  1.57$\pm$0.11 &   66.8$\pm$  2.0 &  1.86$\pm$0.07 & 69.9$\pm$  1.1  &   1.62$\pm$0.05 &  61.8$\pm$  0.9    \\
64	&   2007417656322648576 & 1.21   & 2.64$\pm$0.15 &  63.2$\pm$  1.7   &  1.22$\pm$0.14 &   92.8$\pm$  3.3 &  1.91$\pm$0.13 & 69.0$\pm$  1.9  &   1.94$\pm$0.03 &  61.8$\pm$  0.4    \\
65	&   2007423531838164480 & 1.39   &       -       &      -  	   &  	     - 	    &        -         &  2.05$\pm$0.08 & 71.0$\pm$  1.1  &   1.72$\pm$0.02 &  63.7$\pm$  0.4    \\
66*$^{a}$	&   2007423527528174592 & 1.36   & 2.17$\pm$0.18 &  58.3$\pm$  2.3   &  2.22$\pm$0.21 &   69.5$\pm$  2.7 &  2.26$\pm$0.10 & 69.3$\pm$  1.2  &   2.15$\pm$0.09 &  64.4$\pm$  1.2    \\
67$^{a}$	&   2007423841075526528 & 0.85   & 1.29$\pm$0.07 &  70.2$\pm$  1.6   &  1.43$\pm$0.47 &   70.1$\pm$  9.4 &  1.57$\pm$0.18 & 74.7$\pm$  3.4  &   1.13$\pm$0.15 &  60.5$\pm$  3.8    \\
68	&   2007436794696806016 & 0.44   & 1.70$\pm$0.35 &  41.5$\pm$  5.7   &  0.71$\pm$0.01 &   92.3$\pm$  0.6 &  0.88$\pm$0.09 & 74.2$\pm$  2.8  &   1.16$\pm$0.27 &  47.2$\pm$  6.8    \\
69$^{a}$	&   2007435695185247104 & 0.29   &       -       &      - 	   &  1.48$\pm$0.24 &   55.0$\pm$  4.6 &  1.09$\pm$0.09 & 66.7$\pm$  2.4  &   0.60$\pm$0.04 &  31.0$\pm$  1.9    \\
70$^{a}$	&   2007435729544974976 & 0.59   &      -        &      -  	   &  	  -   	    &        -         &  0.86$\pm$0.12 & 98.7$\pm$  4.0  &   1.33$\pm$0.54 &  77.1$\pm$ 11.6    \\
71$^{a}$	&   2007417862481060096 & 1.33   &       -       &      -            &  4.27$\pm$0.26 &   52.6$\pm$  1.7 &  2.19$\pm$0.12 & 65.8$\pm$  1.6  &   2.19$\pm$0.25 &  67.2$\pm$  3.3    \\
72$^{a}$	&   2007435695185239168 & 1.00   & 4.87$\pm$0.31 &  89.9$\pm$  1.8   &        -       &          -       &  1.22$\pm$0.25 & 75.5$\pm$  5.9  &   1.88$\pm$0.41 &  72.8$\pm$  6.2    \\
\enddata
\tablecomments{ID marked with the asterisk symbol (*) shows the member star of the cluster and ID with the symbol ($^{a}$) are stars with the value of extinction and distance from Anders et al. (2022).}
\end{deluxetable*}

\subsection{Distribution of Polarization Degree and Polarization Angle} 
\begin{figure}[h]
 \centering
 \includegraphics[width=0.98\columnwidth]{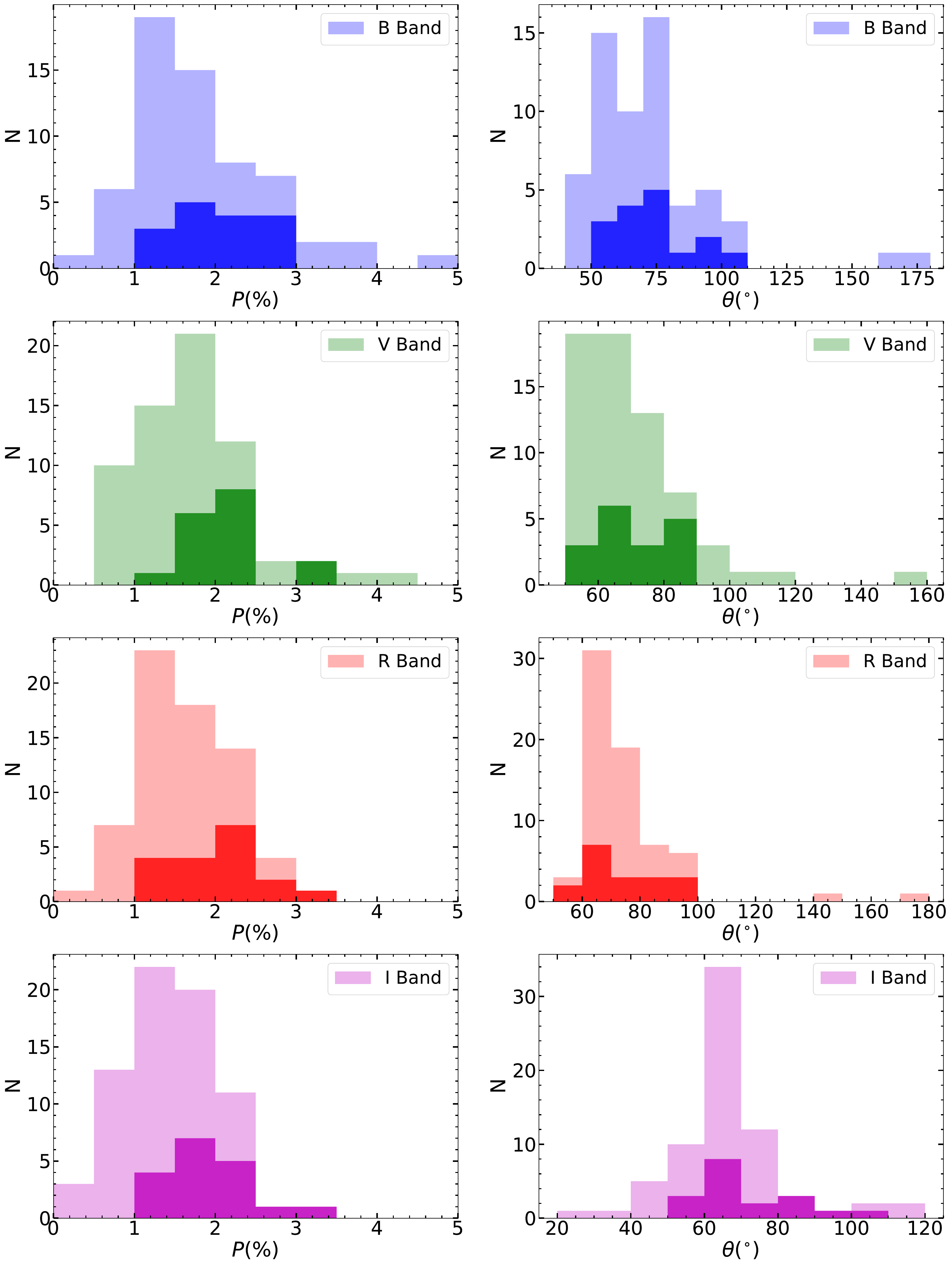}
\caption{The distributions of $P$ (in left panel) and $\theta$ (in right panel) in $B$, $V$, $R$, and $I$ bands. Distributions for all observed and member stars are plotted with the light and dark colors, respectively, in all bands.}
\label{fig:hist_p_the}
\end{figure}

The distributions of the polarization degree ($P$) and angle ($\theta$) among observed sources are shown in Figure \ref{fig:hist_p_the} in $B$, $V$, $R$, and $I$ bands. The distribution of member stars is over-plotted in the figure with the dark color in all bands.  

The polarization degree was found to be in the range of 0.33 to 4.87 $\%$ for the $B$ band, 0.64 to 4.27 $\%$ for $V$, 0.48 to 3.14 $\%$ for the $R$, band, and 0.29 to 3.01 $\%$ for $I$ bands. A single peak distribution of $P$ and $\theta$ was found in $B$, $V$, $R$, and $I$ bands.
The variation of $\theta$ is found as 41$\degr$.5 to 173$\degr$.8 in $B$, 52$\degr$.6 to 156$\degr$.2 in $V$, 58$\degr$.6 to 179$\degr$.8 in $R$, and 29$\degr$.4 to 119$\degr$.8 in $I$ bands. For the majority of stars, the polarization degree lies between 1-2.5$\%$ in all four bands, and the angle lies in the range of 50$\degr$-80$\degr$. 
The mean and standard deviation of $P$ and $\theta$ are derived by fitting the Gaussian function to their distribution. Thus, the average value of $P$ for all observed stars is found as 1.55$\pm$0.64 $\%$, 1.61$\pm$0.61 $\%$, 1.58$\pm$0.62 $\%$, and 1.44$\pm$0.62 $\%$ for $B$, $V$, $R$, and $I$ bands, respectively, and $\theta$ is found as 66$\degr$.2$\pm$16$\degr$.5 in $B$, 64$\degr$.7$\pm$11$\degr$.7 in $V$, 68$\degr$.2$\pm$6$\degr$.6 in $R$, 65$\degr$.4$\pm$6$\degr$.8 in $I$ bands, where the error corresponds to the standard deviation. A polarization variation from 1.3$\%$ to 3$\%$ with a similar position angle was also found by \citet{1976AJ.....81..970M}.

\subsection{Magnetic Fields Projected on the POS from Polarization Vectors}
Polarization angles of starlight polarization are usually used to infer the magnetic field orientation projected on the plane of the sky (POS), over both large and small spatial scales\citep{1970MmRAS..74..139M, 1976MNRAS.177..499A, 1996ApJ...462..316H}. Here, we present the projected magnetic fields of star cluster using starlight polarization.
 
\begin{figure*}[h]
    \centering
\subfigure[$B-$band]{\includegraphics[width=0.45\textwidth,trim={1.5cm 1.5cm 2.0cm 1.0cm}]{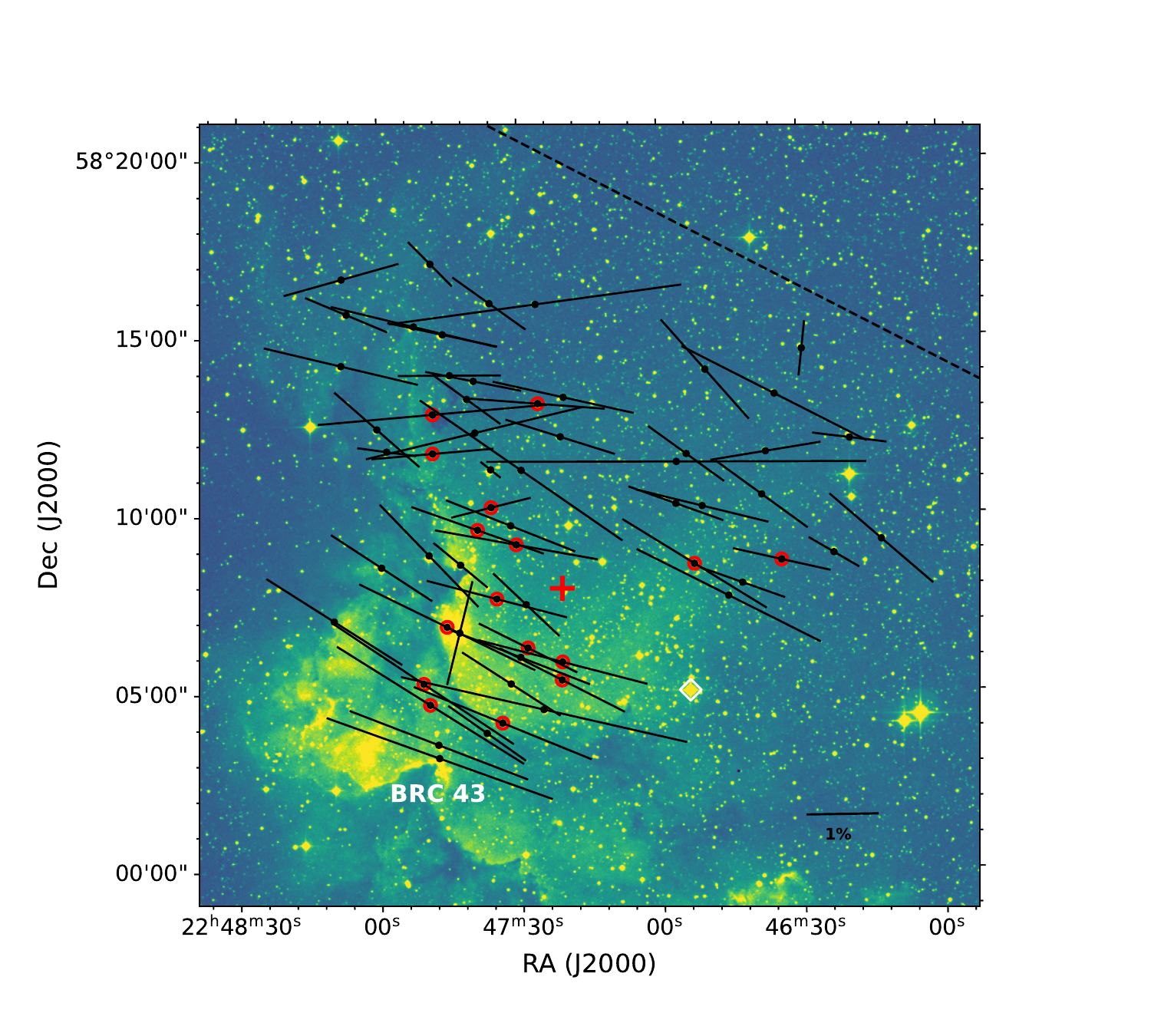}\label{fig:dss_B}}
\subfigure[$V-$band]{\includegraphics[width=0.45\textwidth,trim={1.5cm 1.5cm 2.0cm 1.0cm}]{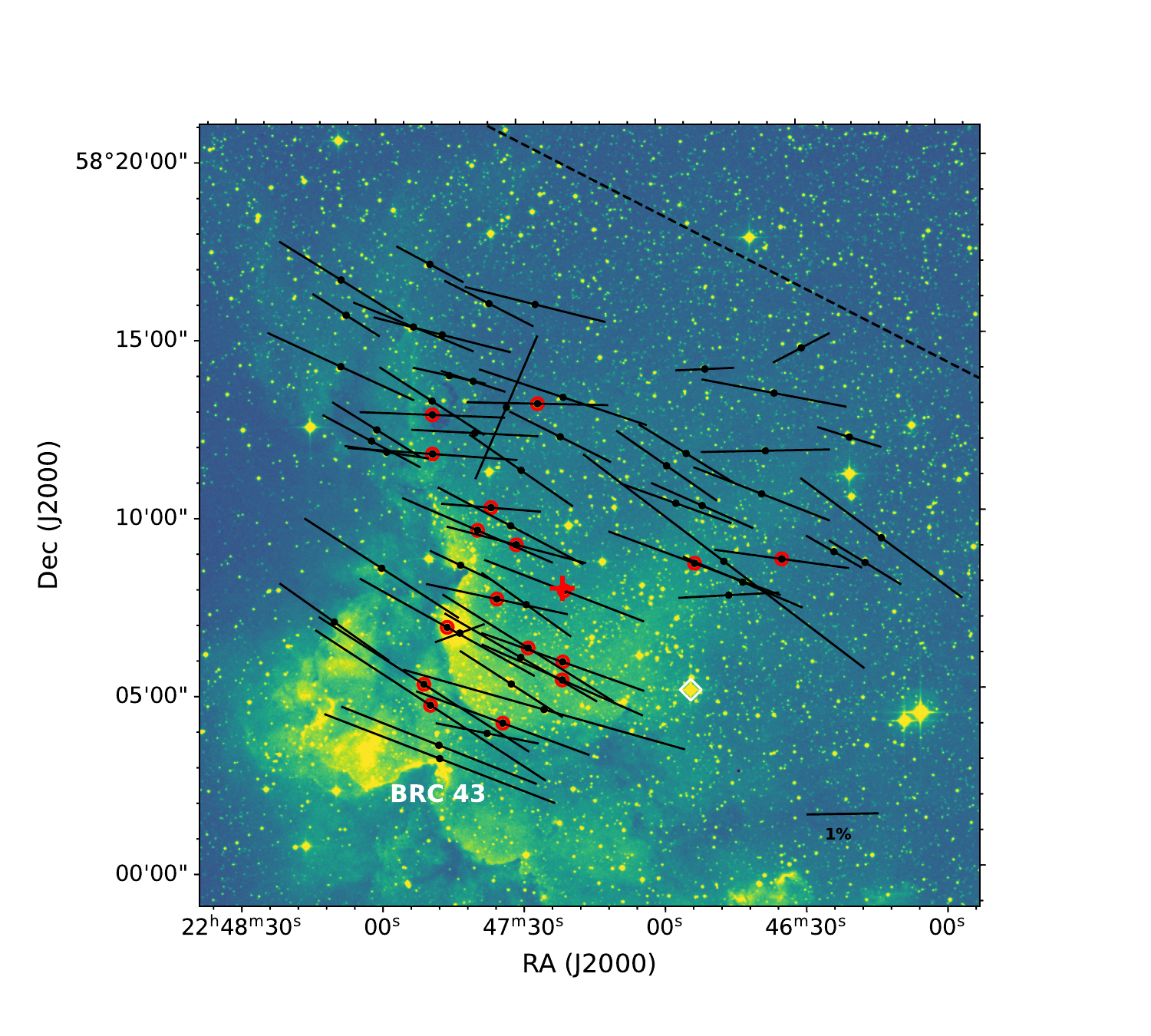}\label{fig:dss_V}}
  \subfigure[$R-$band]{\includegraphics[width=0.45\textwidth,trim={1.5cm 1.5cm 2.0cm 1.0cm}]{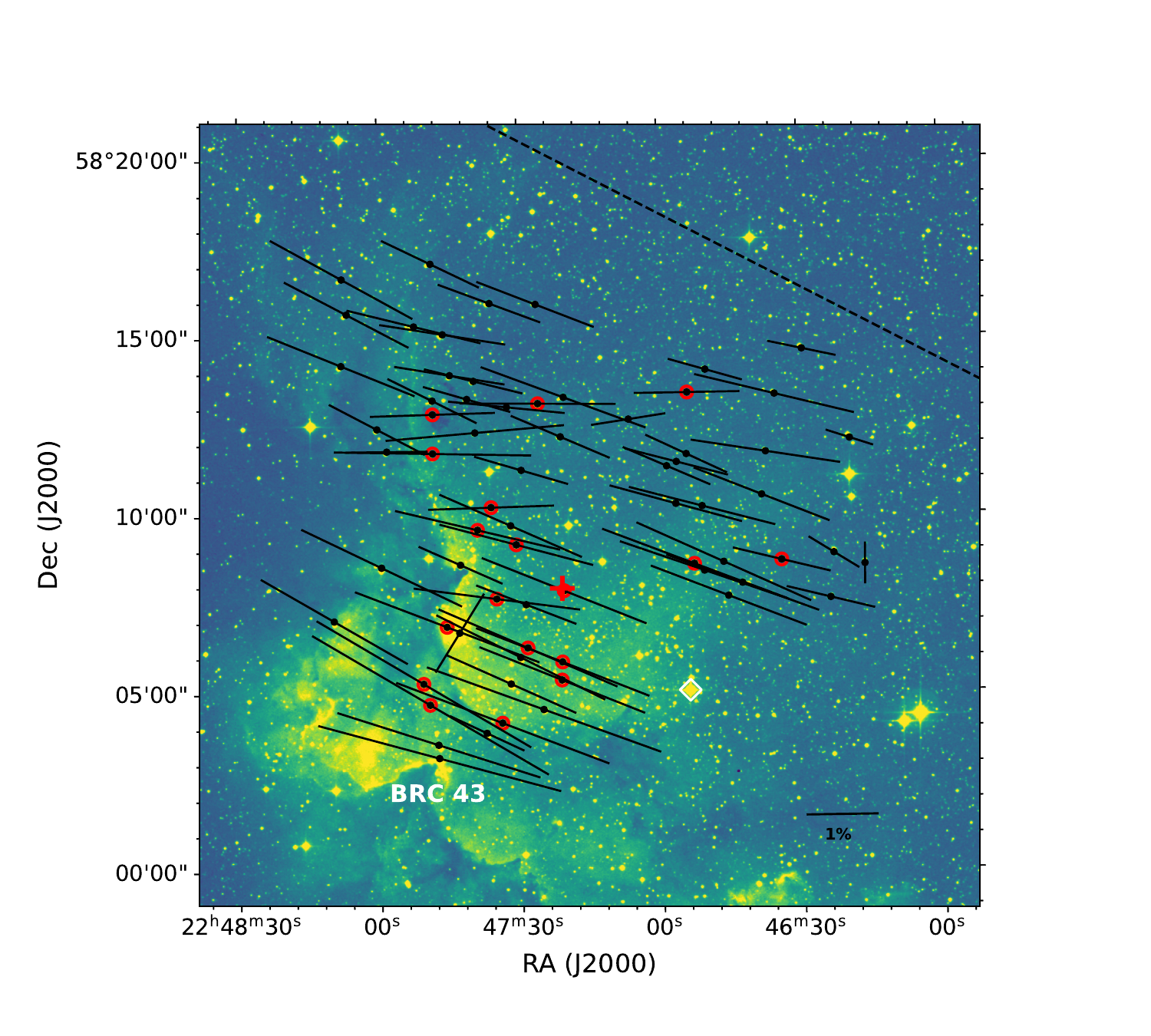}\label{fig:dss_R}} 
  \subfigure[$I-$band]{\includegraphics[width=0.45\textwidth,trim={1.5cm 1.5cm 2.0cm 1.0cm}]{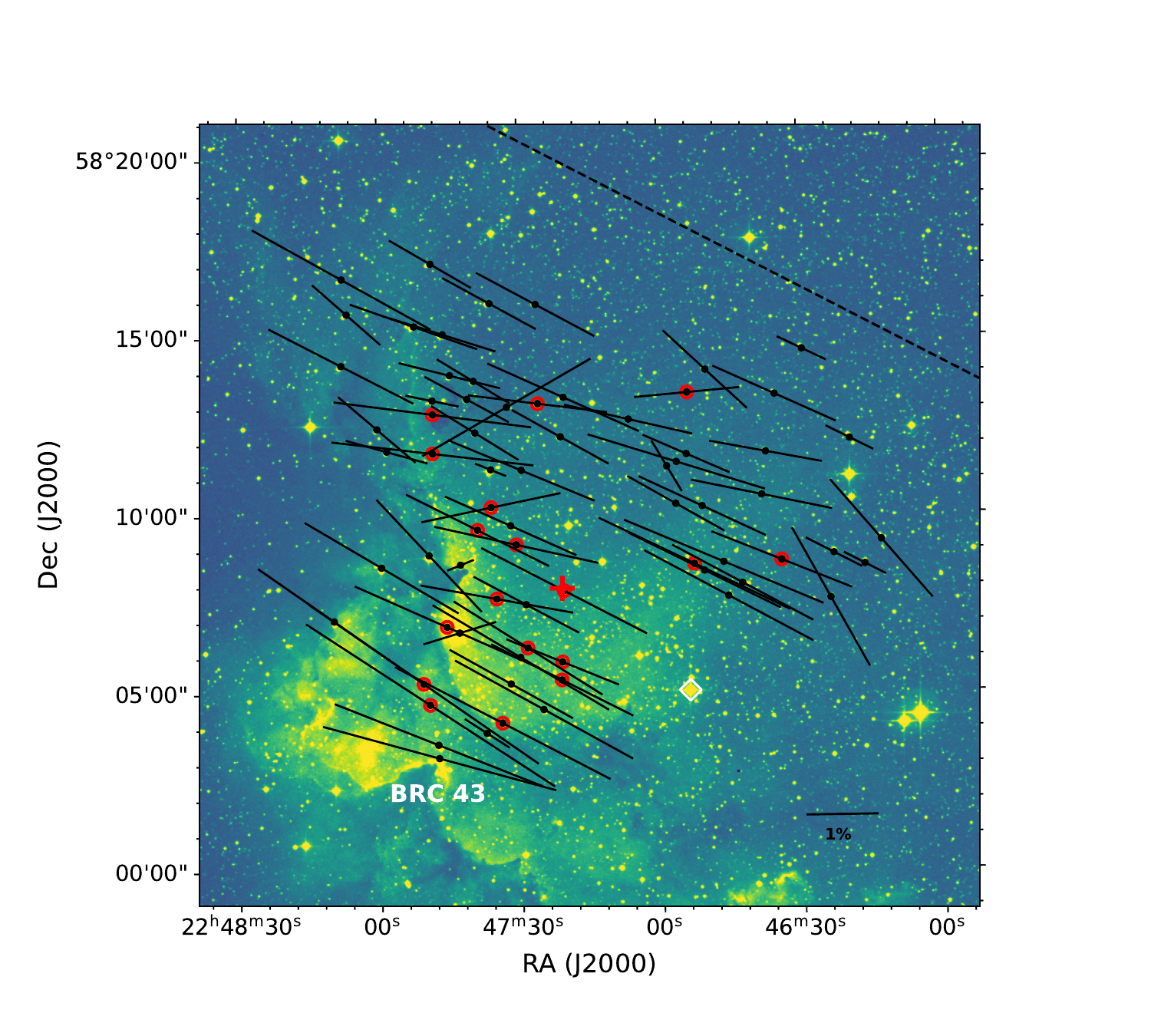}\label{fig:dss_I}}
    \caption{Polarization vectors are over-plotted over the DSS image ($22{^\prime}\times22{^\prime}$) of the observed region. The length of the polarization vectors corresponds to the polarization value, whereas their orientation reveals the POS magnetic fields along the line of sight toward the stars. Member stars are marked with red circles. The dotted line shows the orientation of the projection of GP at the region. The plus symbol (`+') shows the centre of the cluster NGC 7380. The magnetic fields toward these stars are essentially parallel to the Galactic magnetic field.}
    \label{fig:dss}
\end{figure*}

Figure \ref{fig:dss} presents polarization vectors overplotted on the observed region's digital sky survey (DSS)-2 red band image. The length of the vector is a measure of the degree of polarization with the reference length of 1$\%$ polarization at the bottom right. The tilt of these vectors denotes the position angle, which is measured from the north, increasing towards the east. The dotted line in the figure shows the orientation of the projection of the Galactic Plane (GP) in the cluster region ($\theta_{GP} \sim 63\degr$).  
Polarization vectors are nearly parallel to GP's direction, implying that the magnetic fields toward the stars are parallel to the Galactic magnetic field. In $V-$ band, polarization vectors of the stars \#24, \#32, and \#50 are significantly deviated from GP ($|\theta-\theta_{GP}| > 45\degr$). Stars \#24 and \#50  have smaller lengths of polarization vectors relative to the general distribution of polarization vectors. Additionally, it is apparent from the figure that the length of polarization vectors is generally larger around the visible dense dust structure region. Star \#24 is the only star with low polarization and large deviation from GP, despite being embedded in the visible dust cloud. Besides this dense region, polarization vectors are smaller except for one star (\#71), showing substantial polarization. Overall, very few sources have shown uncommon behaviour.

\subsection{Distribution of Dust Extinction in the Region} \label{sec:ptheta}
\begin{figure}
    \centering
\includegraphics[width=1.1\columnwidth,trim={1.4cm 1.0cm 0.2cm 0.0cm}]{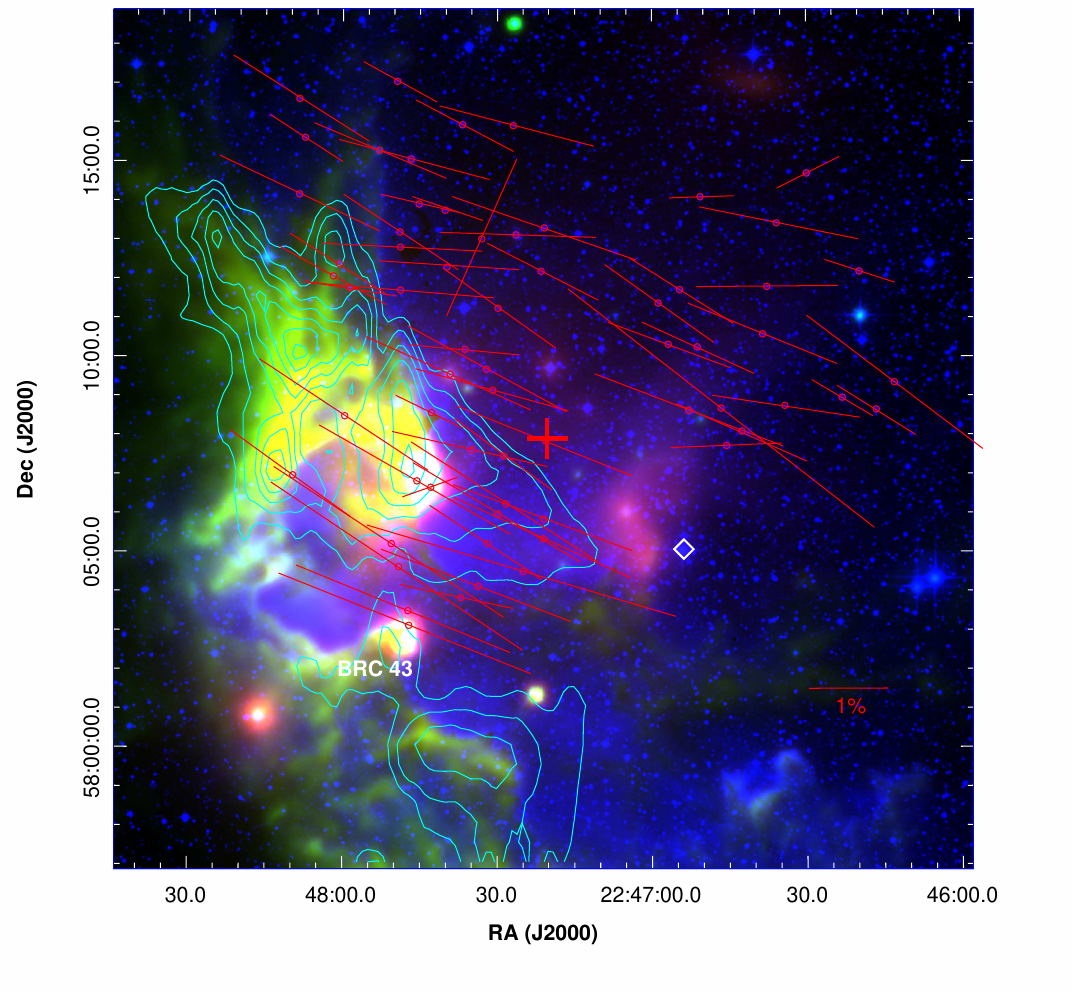}  
  \caption{The colour composite image of the observed region of the cluster NGC 7380 (Red: $WISE-W4$, Green: $WISE-W3$, and Blue: $DSS2-R$). Cyan contours represent the distribution of the dust extinction in the region. Polarization vectors in $V-$ band are overplotted similar to Figure \ref{fig:dss_V}.}
    \label{fig:rgb_image}
\end{figure}

In Figure \ref{fig:rgb_image}, we have shown a three-colour composite (red: $WISE-W4$, green: $WISE-W3$ and blue: $DSS2-R$) image of the region in which $V-$ band polarization vectors are overplotted. $WISE-W4$ (22 $\mu m$) band contains emissions from heated tiny dust grains, mostly near massive stars and embedded young stars. $WISE-W3$ (12 $\mu m$) band is often referred to as the ``Polyaromatic Hydrocarbon (PAH)" $WISE$ band as it includes the wavelengths corresponding to PAH features. PAH is abundant in active star-forming regions and is produced in cold molecular clouds \citep{sand2010}. The PAHs grow on the dust grains, but near the high radiation field of the diffused ISM, these are easily destroyed. In Figure \ref{fig:rgb_image}, we notice that most of the PAH emission is concentrated toward the south-east periphery of the H{\sc ii} region, whereas near cluster, massive star and western side of the region lack PAH emission. The higher values of polarization toward the south-east region very well correlate with the high concentration of PAH emission seen in Figure \ref{fig:rgb_image} and the $^{12}CO$ and $^{13}CO$ line emission found by \citet{2022ApJ...928...17S}. 
The absence of the PAH feature near the massive star indicates the destruction of the PAH due to the high radiation field.
The $WISE$ 22 $\micron$ emission is present near the massive star and embedded young star in the region. 
   
The dust content in the region can be probed via extinction measurement, a tracer for the density distribution of molecular hydrogen in the star-forming regions. The colour excess in the infrared ($IR$) wavelengths is used to measure the extinction \citep{1994ASSL..190..473L}. We generated the extinction map for the NGC 7380 region using the near-infrared ($NIR$) catalogue of $2MASS$ and UKIRT Infrared Deep Sky Survey ($UKIDSS$) data and the following method discussed in \citet{2014MNRAS.443.1614P, 2020ApJ...905...61P}. We divided the region into several small cells and looked for the $(H - K)$ colours of the 20 nearest stars to the cell centre. We computed the colour excess $E = (H - K) - (H - K)_{0}$, where $(H - K)$ is the observed median colour of a cell and $(H - K)_{0}$ is the intrinsic median colour of unreddened stars. Using the equation $A_{K} = 1.82 E$ \citep{2007ApJ...663.1069F}, we calculated the extinction in the $K$-band within each cell and further converted it to the $A_{V}$ using the relation $A_{V}$ = 12.82 $\times$ $A_{K}$.
The extinction map for the region is plotted in Figure \ref{fig:rgb_image} as cyan contours. The contours start at $\sim$ 35$\%$ of the peak value with an increment of  $\sim$ 10$\%$. The extinction map's resolution depends on the stars' surface density. It is higher for regions with high surface density, and for low surface density regions, the resolution is relatively lower. 
The extinction map indicates a higher concentration of dust toward the eastern side of the cluster, whereas there is a deficit of dust toward the west and north of the cluster. The extinction map nicely overlaps with the PAH emission features revealed by the $WISE$-12 micron images. Our dust extinction map also has a similar distribution to the molecular gas obtained through the CO observations of the region by \citet{2022ApJ...928...17S}.

\subsection{Properties of Interstellar Polarization} \label{sec:ser}
The interstellar polarization or polarization of starlight can be obtained from multi-wavelength polarimetric observations. An empirical relation between the interstellar polarization and wavelength was found to be followed, and the relation was established by \citet{1975ApJ...196..261S}. The relation for the variation of interstellar polarization with the wavelength is described by
\begin{equation}
P_{\lambda} = P_{\rm max} \times \exp[-K \ln^2(\lambda_{\rm max}/\lambda)],
\label{eq:ser_curve}
\end{equation}
where $P_{\lambda}$ is the polarization at wavelength $\lambda$, and $P_{\rm max}$ and $\lambda_{\rm max}$ are the maximum polarization and wavelength corresponding to it, respectively. The above Serkowski relation represents the ISM polarization in the wavelength range 0.36-1 $\mu m$ with $\lambda_{\rm max}$ in the 0.45 to 0.8 $\mu m$ range.

The parameter $K$ was usually taken as constant of $K=1.15$ in previous studies \citep[e.g.][]{1979AJ.....84..812P, 2007MNRAS.378..881M, 2011MNRAS.411.1418E, 2020AJ....159...99S, 2020AJ....160..256S}.  
However, it was found to be correlated with the $\lambda_{\rm max}$ as $K=(-0.002\pm0.07) + (1.68\pm0.13) \lambda_{\rm max}$ \citep{1980ApJ...235..905W}, $K=(-0.10\pm0.05) + (1.86\pm0.09) \lambda_{\rm max}$ \citep{1982AJ.....87..695W}. Later, \citet{1992ApJ...386..562W} modelled $P_{\lambda}$ in wavelength range 0.36-2 $\mu m$ and found the modification in the parameter $K$ as $K=(0.01\pm0.05) + (1.66\pm0.09)\lambda_{\rm max}$. Physically, the value of $\lambda_{\rm max}$ describes the average size of aligned grains \citep{1980ApJ...235..905W, 1995ApJ...444..293K, 2020ApJ...905..157V}. The parameter $K$ denotes the inverse of the width of the polarization curve; hence, the relation between $K$ and $\lambda_{\rm max}$ reflects the reduced width with the increase in $\lambda_{\rm max}$, which physically signifies the narrowing size distribution of aligned grains as the minimum size of aligned grains increases \citep{2020ApJ...896...44L}. 

Here we have adopted the $K=1.66\times \lambda_{\rm max}+0.01$ relationship \citep{1992ApJ...386..562W, 2004A&A...419..965M, 2008MNRAS.391..447F, 2010A&A...513A..75O, 2012MNRAS.419.2587E, 2018RMxAA..54..293V}. The error in polarization is also considered in the fitting of Serkowski relation for each star and the value of $P_{\rm max}$, $\lambda_{\rm max}$, and $\sigma_{1}$ (unit weight error of fit) are extracted. The best-fit values of $P_{\rm max}$, $\lambda_{\rm max}$, and $\sigma_{1}$ for the observed stars are given in Table \ref{tab:serkowski}. We also show the normalized angle dispersion in degree ($\bar{\epsilon}$), obtained using the dispersion of the position angles in all bands divided by the mean error of angles. Stars, for which errors in $P_{\rm max}$ and/or $\lambda_{\rm max}$ are greater than $2\sigma$ (i.e. relative error $>$ 50$\%$), are not included in the table. 

\startlongtable
\begin{deluxetable}{cccccc}
\tablecaption{The $P_{\rm max}$, $\lambda_{\rm max}$, $\sigma_{1}$, and $\bar{\epsilon}$ for stars in the region of the cluster NGC 7380.}
\label{tab:serkowski}
\tablehead{
\colhead{S. No.} & \colhead{$P_{\rm max}$ ($\%$)} & \colhead{$\lambda_{\rm max}$ ($\mu m$)} & \colhead{$\sigma_{1}$} & \colhead{$\bar{\epsilon}$ ($\degr$)} & \colhead{$A_{V}$ (mag)}
}
\startdata
1    &  2.04$\pm$0.10	&   0.54$\pm$0.07   &  0.6 & 0.80  & -\\
2    &  2.10$\pm$0.14	&   0.53$\pm$0.06   &  1.0 & 2.77   & 0.919\\
3*    &  2.98$\pm$0.06	&   0.59$\pm$0.03   &  0.4 & 1.78   & 1.564\\
4*    &  3.23$\pm$0.06	&   0.57$\pm$0.04   &  0.5 & 0.98   & 0.949\\
5    &  2.54$\pm$0.03	&   0.57$\pm$0.01   &  0.1 & 0.93  &  1.429\\
6    &  3.20$\pm$0.15	&   0.45$\pm$0.04   &  0.6 & 7.71  & - \\
7*    &  2.49$\pm$0.03	&   0.48$\pm$0.02   &  0.2 & 5.28  & - \\
9*    &  1.98$\pm$0.12	&   0.52$\pm$0.07   &  0.6 & 4.61  &  1.487 \\
10*   &  1.79$\pm$0.06	&   0.58$\pm$0.09   &  0.5 & 1.51  & 1.506 \\
11   &  1.85$\pm$0.12	&   0.47$\pm$0.05   &  0.6 & 2.04  &  1.246\\
12*   &  1.80$\pm$0.05	&   0.63$\pm$0.06   &  0.6 & 0.76  &  1.681\\
13   &  1.33$\pm$0.01	&   0.54$\pm$0.05   &  0.2 & 2.87  &  1.705\\
14*   &  1.85$\pm$0.08	&   0.67$\pm$0.08   &  1.4 & 2.77  & 1.275 \\
15   &  2.17$\pm$0.01	&   0.66$\pm$0.01   &  0.2 & 4.04  &  1.326\\
16   &  1.53$\pm$0.06	&   0.63$\pm$0.07   &  0.5 & 3.26   & -\\
17*   &  2.66$\pm$0.09	&   0.77$\pm$0.05   &  0.9 & 1.61  & 1.827 \\
19*   &  2.06$\pm$0.01	&   0.59$\pm$0.01   &  0.0 & 2.13  &  1.278\\
20*   &  2.13$\pm$0.04	&   0.61$\pm$0.05   &  1.0 & 0.99  &  1.020\\
21*   &  2.14$\pm$0.25	&   0.44$\pm$0.09   &  0.8 & 1.67   & 1.484 \\
25*   &  1.43$\pm$0.06	&   0.74$\pm$0.03   &  0.6 & 27.37  &  1.816\\
27   &  2.17$\pm$0.12	&   0.48$\pm$0.06   &  0.4 & 1.35   & 1.563\\
28   &  1.39$\pm$0.08	&   0.46$\pm$0.05   &  0.2 & 2.29  &  0.836\\
29*   &  1.77$\pm$0.05	&   0.51$\pm$0.04   &  0.4 & 1.65  &  1.707 \\
32   &  1.94$\pm$0.20	&   0.80$\pm$0.25   &  0.9 & 4.67  &  2.761\\
33   &  1.25$\pm$0.02	&   0.60$\pm$0.03   &  0.1 & 1.88  &  2.181 \\
34   &  1.14$\pm$0.17	&   0.55$\pm$0.23   &  0.6 & 1.37   & 1.491 \\
35   &  1.06$\pm$0.01	&   0.59$\pm$0.03   &  0.2 & 3.24   & 0.927 \\
39*   &  2.26$\pm$0.04	&   0.63$\pm$0.06   &  0.7 & 2.95   & 2.011\\
40   &  1.68$\pm$0.14	&   0.51$\pm$0.11   &  0.6 & 1.32  &  2.116\\
41   &  0.99$\pm$0.03	&   0.68$\pm$0.05   &  0.5 & 3.88  &  0.480\\
42   &  1.91$\pm$0.07	&   0.69$\pm$0.04   &  0.3 & 5.67   & -\\
43   &  1.29$\pm$0.30	&   0.74$\pm$0.28   &  1.2 & 1.86   & 1.128\\
44   &  1.96$\pm$0.04	&   0.51$\pm$0.02   &  0.1 & 2.24  &  1.359\\
45   &  1.28$\pm$0.03	&   0.51$\pm$0.07   &  0.4 & 2.66  &  1.446 \\
46   &  0.97$\pm$0.04	&   0.56$\pm$0.07   &  0.3 & 1.60  &  1.823\\
48   &  1.47$\pm$0.05	&   0.51$\pm$0.06   &  0.6 & 1.93  &  1.850\\
49   &  1.23$\pm$0.43	&   0.87$\pm$0.19   &  1.2 & 1.85  & 1.269 	 \\
50   &  0.77$\pm$0.01	&   0.58$\pm$0.09   &  0.3 & 8.98   & 0.414 \\
54   &  1.78$\pm$0.04	&   0.51$\pm$0.03   &  0.3 & 2.48  &  1.035\\
55   &  0.80$\pm$0.04	&   0.44$\pm$0.06   &  0.1 & 0.47  & 1.215 \\
58   &  1.87$\pm$0.20	&   0.57$\pm$0.11   &  0.6 & 0.74  &  -\\
59*   &  1.54$\pm$0.11	&   0.88$\pm$0.18   &  0.9 & 0.91  &  1.461\\
61   &  1.28$\pm$0.25	&   0.40$\pm$0.15   &  0.3 & 1.16  &  0.564\\
62   &  1.81$\pm$0.04	&   0.53$\pm$0.03   &  1.1 & 5.32  & 0.943 \\
63   &  1.72$\pm$0.07	&   0.69$\pm$0.08   &  0.6 & 1.65   & 1.165\\
64   &  2.12$\pm$0.35	&   0.58$\pm$0.19   &  2.1 & 6.84  &  -\\
66*   &  2.32$\pm$0.03	&   0.60$\pm$0.02   &  0.2 & 2.47  &  1.134\\
67   &  1.39$\pm$0.08	&   0.58$\pm$0.06   &  0.4 & 1.14   & 1.103\\
68   &  0.95$\pm$0.42	&   0.86$\pm$0.24   &  1.5 & 5.18  &  -\\
\enddata
\tablecomments{ID marked with the asterisk symbol (*) shows the cluster's members.}
\end{deluxetable}

Table \ref{tab:serkowski} shows that the value of $P_{\rm max}$ in the observed region varies from 0.77-3.23$\%$, whereas $\lambda_{\rm max}$ varies in the range of 0.40-0.88 $\mu m$. A Gaussian fit to the distribution of $P_{\rm max}$ and $\lambda_{\rm max}$ was performed, and the average value and standard deviation for these parameters were extracted. Towards the observed region of the cluster NGC 7380, the average values of $P_{\rm max}$ and $\lambda_{\rm max}$ are found as 1.71$\pm$0.60 $\%$ and 0.55$\pm$0.07 $\mu m$. The value of $\sigma_{1}$ defines a limit for differentiating between intrinsic and interstellar polarization. The larger value of $\sigma_{1}$ ($>$1.6) shows the intrinsic nature of polarization as the relation was not well followed. The lower value of $\lambda_{\rm max}$ also indicates the intrinsic polarization in sources. 

We note that, for the general diffuse ISM, $\lambda_{\rm max}$ lies in the range of 0.45-0.8 $\mu m$ with the average value of 0.55 $\mu m$. We also found a similar range of $\lambda_{\rm max}$ in observed stars. However, there are very few stars (\#32, \#49, \#59, \#68), that show $\lambda_{\rm max}$ $\geq$ 0.8 $\mu m$, but the error associated with the value of $\lambda_{\rm max}$ is higher, that gives less weight-age to these values.
Based on the $\sigma_{1}$ criteria, only one star (\#64) has indicated intrinsic polarization. No stars in the region are found with the significantly lower value of $\lambda_{\rm max}$. Additionally, most stars have not shown any significant variation of position angle with the wavelength. However, there are few stars (\#6, \#7, \#25, \#42, \#50, \#62, \#64, \#68) which are showing dispersion ($\bar{\epsilon} > 5$) of position angle in different bands, of which star \#25 has shown a very high $\bar{\epsilon}$. These stars are the probable candidates for an intrinsic component of polarization other than interstellar. Star \#50 was also differentiated with a shifted $V-$ band position angle from GP and a smaller degree of polarization. This star is a foreground star, so the smaller degree of polarization could be due to the lower column density of dust grains. However, the dispersion of angles in different bands and the different angles from GP could be due to the possibility of intrinsic polarization of this star. Star \#25 is near the cluster distance, so it was expected to have a high polarization, but due to another component of polarization, this star seems to be depolarized.  
\citet{2016MNRAS.456.2505L} have found  57 stars as variables in the cluster NGC 7380. We found that 9 stars (\#1, \#2, \#5, \#10, \#11, \#30, \#37, \#64, \#65) are common in both studies. Despite being variable, their polarization has not shown any intrinsic component of polarization except star \#64. In the polarization study of DH Cep, \citet{2023arXiv230517259Q} observed nearly a dozen field stars for deriving the interstellar component and hence determined DH Cep intrinsic polarization. Five stars (\#3, \#7, \#9, \#10, \#29) from our observed stars are found in their data with the polarization value. The polarization and position angle for two stars (\#3, \#7) exactly match in both works in all bands, and for three stars (\#9, \#10, \#29), the difference in the degree of polarization is less than 0.2 per cent.

To extract values of $P_{\rm max}$ and $\lambda_{\rm max}$ for the interstellar medium towards the line of sight of the cluster NGC 7380, we have not considered these probable candidates for intrinsic polarization. Hence, the mean values of $P_{\rm max}$ and $\lambda_{\rm max}$ for ISM are derived as 1.71$\pm$0.57$\%$ and 0.56$\pm$0.07 $\mu m$ with the Gaussian fitting to the distribution.

\subsection{The Stokes ($Q-U$) Plane} \label{sec:qu}
\begin{figure}
    \centering
\includegraphics[width=0.98\columnwidth]{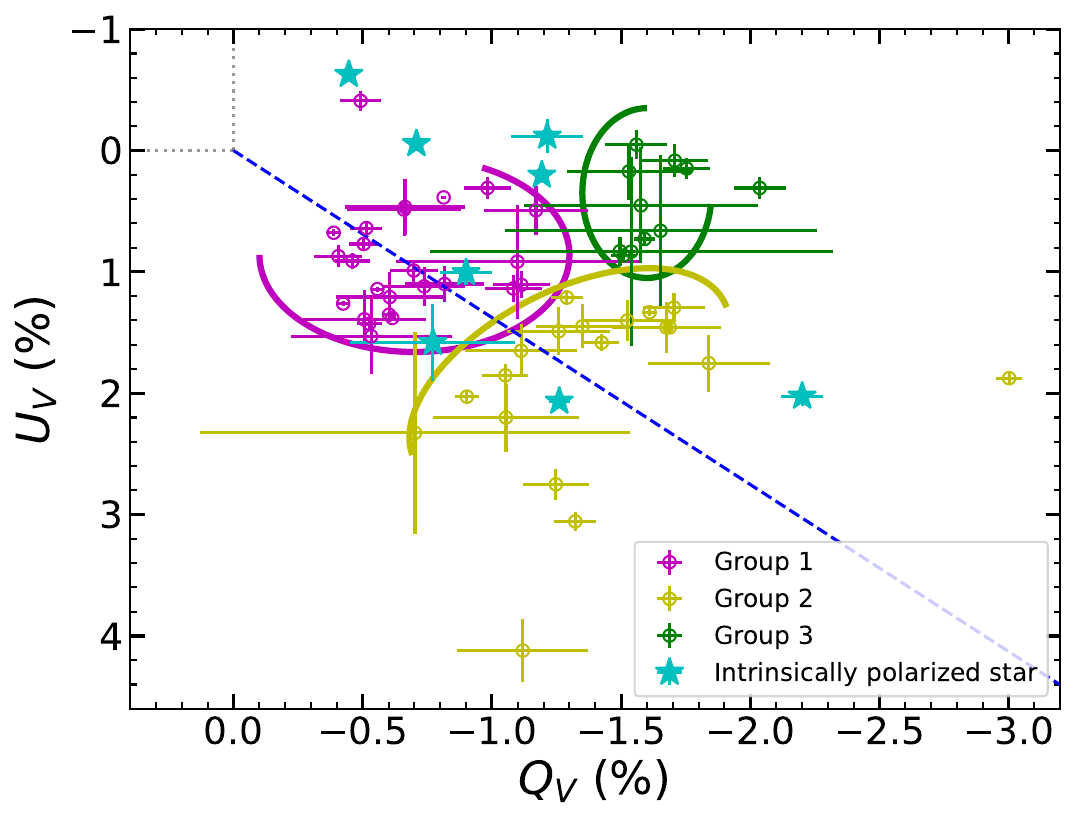} 
\caption{The plot between Stokes parameters $Q$ and $U$ in the $V-$ band. The three groups are shown in three different colors. The figure shows no highly dispersed star (\#32).} 
\label{fig:QU} 
\end{figure}

The interstellar environment's details and interstellar clouds' presence towards the line of sight can be investigated by analysing the Stokes plane.
The origin of ISM polarization is a result of the passage of starlight through aligned columnar dust grains. So, if there are different clouds with different dust distributions in the line of sight, then the ISM-induced polarized starlight will also contain this information. Moreover, the same will be reflected in Stokes's plane. By different distributions of dust, we mean differences in their concentration, size, alignment, etc. There will be a difference in the polarization properties of the starlight that passes through multiple clouds, and it will contain information about the path of those clouds.       
To see the interstellar environment for the region of the cluster NGC 7380, we have shown the Stokes plane of observed sources in Figure \ref{fig:QU}. Stokes plane is a plot between Stokes parameters $Q$ and $U$, which are obtained using $Q=P \cos(2\theta)$ and $U=P \sin(2\theta)$. The point (0,0) in the plane denotes the dustless solar neighbourhood, and the blue dotted line from (0,0) shows the direction of GP.

In general, observed sources are found to be distributed along the line of GP, specifying that no other strong component of the magnetic field is present. 
However, three different groupings of stars appear in the Stokes plane. Two groups are emerging along the GP line, while one significantly differs from the GP direction. We name the group parallel to GP but close to the solar neighbourhood as Group 1, another in the GP line as Group 2, and far from the GP line as Group 3. We have drawn three curves with magenta, yellow, and green colours separating these Groups 1, 2 and 3. 
These three groups are also found to be very well separated in the $P-\theta$ diagram. Group 1 has low polarization, Group 2 has high polarization, and Group 3 stars have polarization between Groups 1 and 2 but witPh a higher position angle. The same was also confirmed by seeing the polarization and position angle histogram for these groups. 
In the sky plane of the observed region, we noticed that group 2 stars are mainly concentrated towards the southeast region. In contrast, other group stars are diverse in other remaining parts of the region.

\subsection{Variation of Polarization with Distance} \label{sec:p_dist}
With precise information on the distance of sources, it could have been possible to aim the three-dimensional structure of dust distribution. Even though, while targeting cluster, we observe many foreground and background sources. With these sources, we can see how polarization properties vary with distance. 
The information regarding the foreground dust concentration towards the line of sight can be extracted using the distribution of polarization with distance. A constant increase of polarization with distance is expected with the uniformly aligned grains towards the direction. With the distance, the starlight travels through more column density of ISM grains. Hence, the polarization will increase if the magnetic field's orientation does not change significantly across the line of sight. However, polarization doesn't need to increase linearly with distance. Generally, high dust concentration is not available throughout. It was found that highly concentrated dust is present mostly in dust layers. 

\begin{figure}
    \centering
\includegraphics[width=0.48\textwidth]{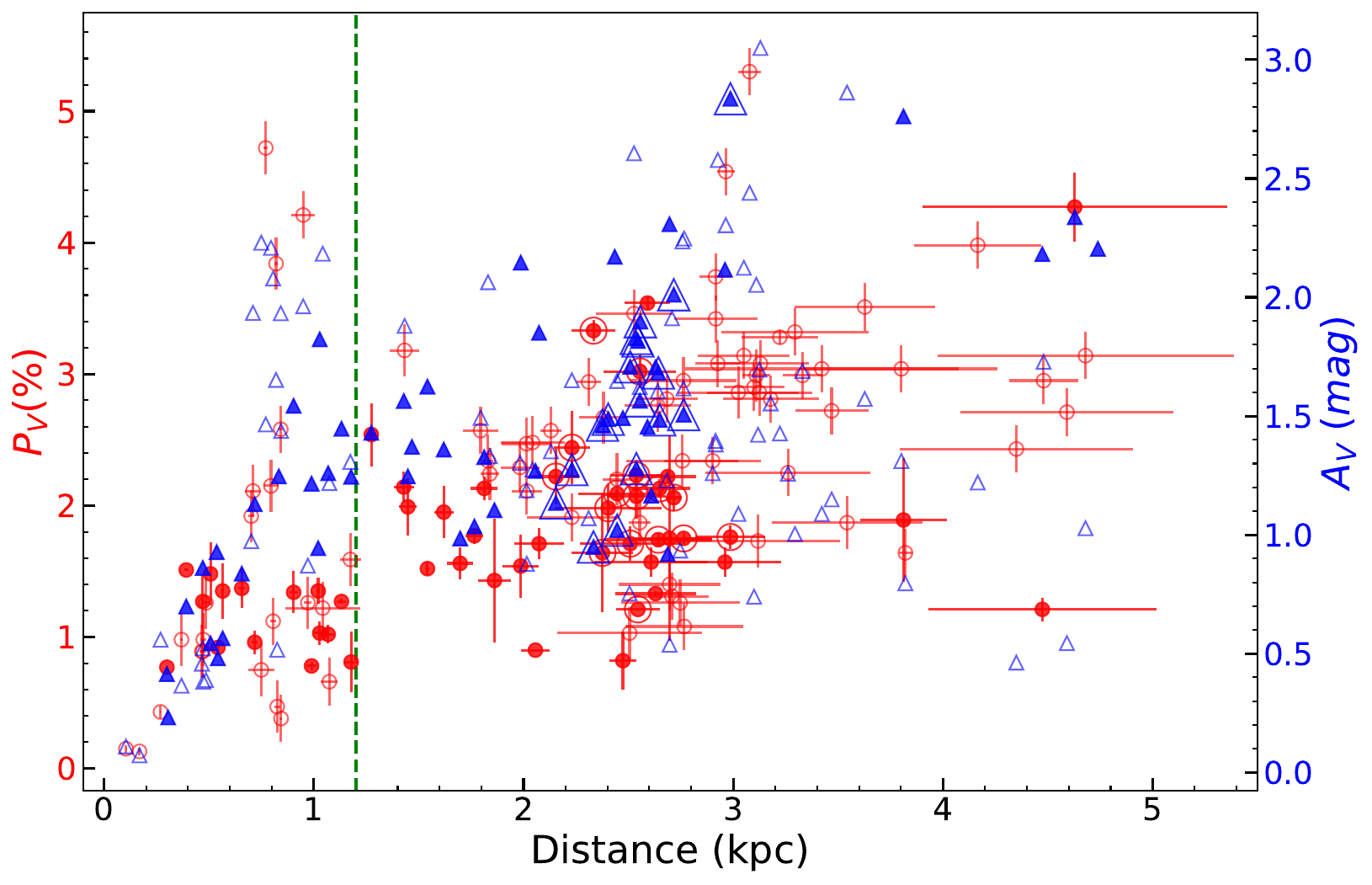}
  \caption{Variation of $P_{V}$ and $A_{V}$ with distance. The red and blue markers are for $P_{V}$ and $A_{V}$, respectively. Filled markers show the observed stars by us, and open markers are for the sources from \citet{2000AJ....119..923H}. The filled marker with an additional marker layer shows the member stars. One background Heiles star (distance $>$ 6 kpc) is left for the plot. The vertical line at 1.2 $kpc$ distance denotes the location of the dust layer.}
    \label{fig:dist_pv}
\end{figure}

To enrich the polarization data at different distances, we have supplemented the data for degree and angle of polarization from \citet{2000AJ....119..923H}. \citet{2000AJ....119..923H} presented an optical polarization catalogue for 9286 stars from the collection of many smaller catalogues. We have taken the polarization data within the $5\degr$ radius of the cluster centre from this catalogue. 
We have matched these sources with the Gaia sources with the cross-match criteria within 1\arcsec. Distances of these sources are taken from \citet{2022A&A...658A..91A}. We are left with 98 useful sources from this catalogue as some sources were found to have $P/\sigma_{P}$ $<$ 2, some were not matched within 1\arcsec, and for some sources, there was no estimation of distance in \citet{2022A&A...658A..91A}. These additional 98 sources' polarization data from \citet{2000AJ....119..923H} is used in our analysis. With the Gaia unique Id of sources, we have verified that no sources are common in observed and \citet{2000AJ....119..923H} data sets. 
We also took extinction data for these 98 sources from \citet{2022A&A...658A..91A} as extinction is a reliable tracer of dust concentration. Following the similar approach as mentioned above, we have shown $P_{V}$ and $A_{V}$ with distance in Figure \ref{fig:dist_pv} for observed and Heiles sources. The error in distance is taken as =(d84 - d16)/2, where d84 and d16 are distances at the 84$^{th}$ and 16$^{th}$ percentiles.
The red scale is for $P_{V}$, and the blue scale is for $A_{V}$. Red-filled circles and empty red circles denote the polarization for observed and Heiles sources, and blue-filled triangles and empty blue triangles show the extinction for observed and Heiles sources, respectively.

Figure \ref{fig:dist_pv} shows that $P_{V}$ and $A_{V}$ increase gradually up to near the distance of $\sim$1.2 $kpc$, then a slight increase in both $P_{V}$ and $A_{V}$ occurs. After this increase, nearly constant values but with a large dispersion are observed. The increase of $P_{V}$ at the distance of $\approx$1.2 $kpc$ is correlated with the increase in $A_{V}$, which is probably due to the existence of a dust layer at this distance towards the line of sight of the cluster NGC 7380. However, we did not notice any large change in the polarization angle, $\theta_{V}$, with distance, indicating that all dust clouds or layers covering the line of sight have a roughly similar magnetic field orientation. 

Moreover, our polarization data show a linear increase of $P_{V}$ with $A_{V}$ up to the extinction $\approx$ 1.2 $mag$ and above this extinction, a large dispersion is noticed. Such a linear increase of the polarization with $A_{V}$ is expected for a homogeneous distribution of dust along the line of sight. In some survey studies (e.g., \citealt{2023AJ....165...87V}), it was seen that for lower extinction, the polarization degree follows this trend; however, for higher extinction, the relation flattens out.

\subsection{Polarizing Efficiency}
The polarization and extinction are correlated but differ from each other as all types of dust grains contribute to the extinction, while some specific types of grains are responsible for polarization. The specification lies in their shape (should be asymmetric), size (larger than cutoff size 0.05 $\mu m$), and composition (silicates type). It was found that the asymmetric larger size and silicate-type grains get aligned in the interstellar medium. All sizes, shapes, and compositions of grains cause extinction. Therefore, the polarizing efficiency is measured from the produced polarization for a given extinction. Polarizing efficiency is the amount of polarization per unit extinction and is a function of magnetic field orientation and its strength, efficiency of alignment, etc. 

\begin{figure*}
    \centering
    \subfigure[]{\includegraphics[width=0.48\textwidth]{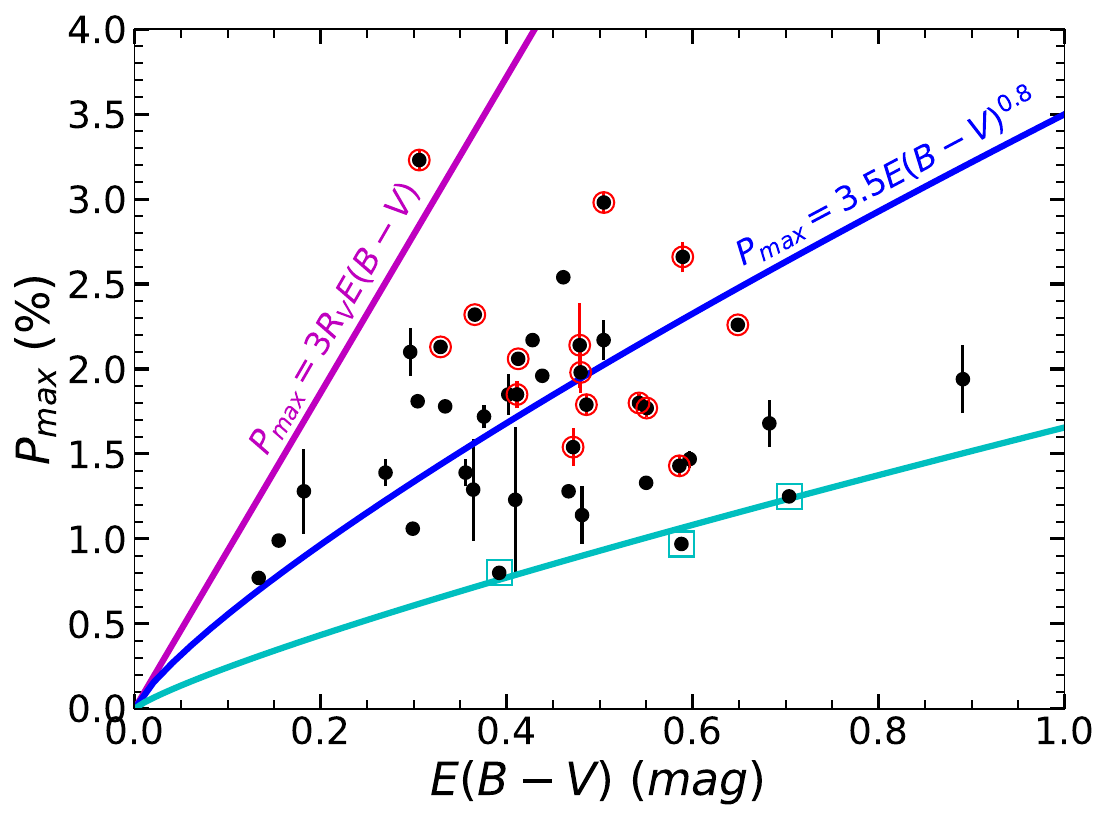}\label{fig:eff1}}
   \subfigure[]{\includegraphics[width=0.48\textwidth]{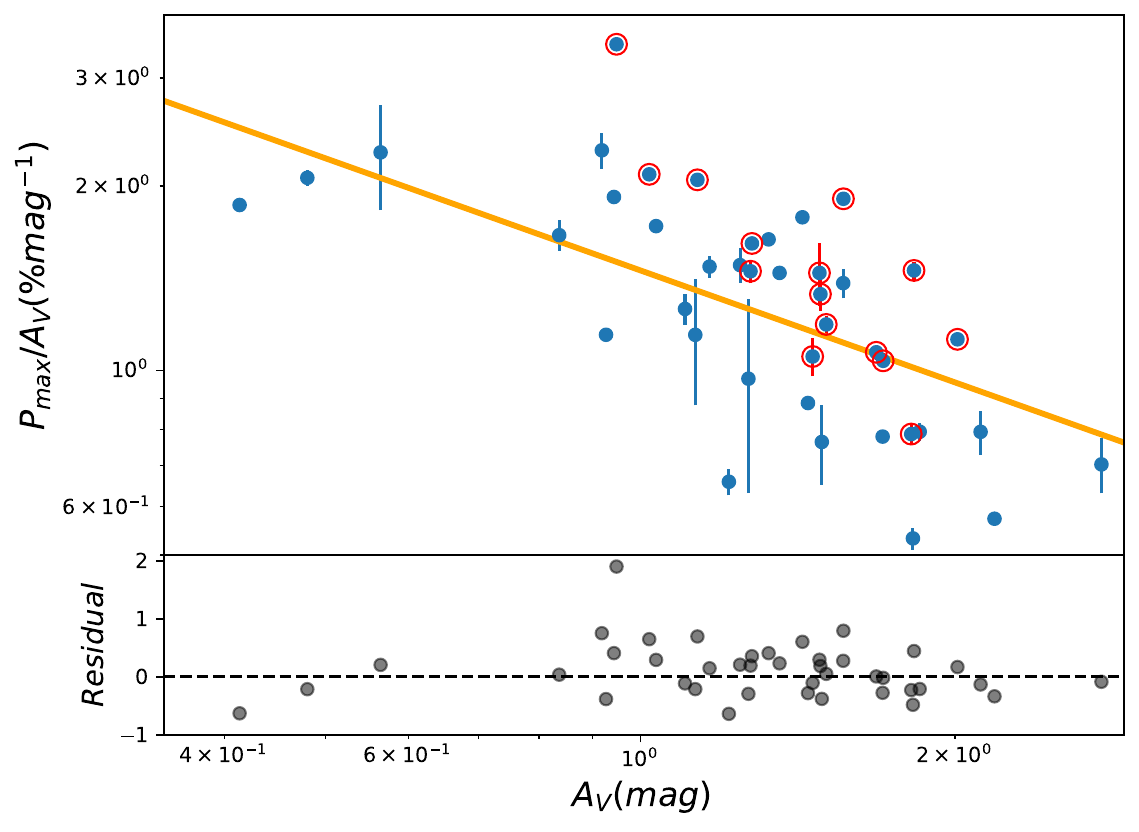}\label{fig:eff2}} 
    \caption{(a) The variation of $P_{\rm max}$ with $E(B-V)$. The magenta, blue, and cyan curves denote the maximum efficiency line, average efficiency for diffuse ISM, and lower bound on efficiency, respectively. (b) $P_{\rm max}/A_{V}$ versus $A_{V}$ in log-log scale along with the residual plot. The marker with the red circle denotes the member stars of the cluster.}  
    \label{fig:eff}
\end{figure*}

We have shown the variation of $P_{\rm max}$ with the reddening [$E(B-V)$] in Figure \ref{fig:eff1}. The reddening was calculated from extinction using the value of 3.1 for total-to-selective extinction [$R_{V}=A_{V}/E(B-V)$], as normal reddening law was found to be followed towards the cluster NGC 7380 \citep{1994A&A...283..963C, 2011AJ....142...71C}. 
It was found in several observational studies that although the increase in polarization with reddening is expected, it can not exceed a particular upper limit. The upper limit for diffuse ISM is given by \citep{1956ApJS....2..389H, 1975ApJ...196..261S},  
\begin{equation}
P_{\rm max} \leq 3 R_{V} \times E(B-V).
\label{eq:upper_limit}
\end{equation}
The magenta line in the figure shows the maximum efficiency line for $R_{V}$ = 3.1. All stars lie below the maximum efficiency line except one star (Id: \#4). This star is a cluster member, but intrinsic polarization is not detected in the star as discussed in Section \ref{sec:ser}. 
According to the relation given by \citet{2002ApJ...564..762F}, the average efficiency for the Galaxy can be given by, 
\begin{equation}
P_{\rm max}=3.5 \times E(B-V)^{0.8}.
\label{eq:av_eff}
\end{equation}
The blue curve shows the average efficiency curve for the Galaxy. The majority of stars are seen to be located close to the average efficiency curve. This indicates that, on average, the polarizing efficiency in the observed region is similar to the general diffuse ISM.  
Additionally, we fitted a power law using three stars (shown by cyan squares) to determine the lower bound on the polarizing efficiency for the region. We derived the relation as,
\begin{equation}
P_{\rm max}=1.66 \times E(B-V)^{0.83},
\label{eq:lower_limit}
\end{equation}
which is shown by the cyan curve in the figure. The observed data points are lying within the lower limit (Equation \ref{eq:lower_limit}) and upper limit (Equation  \ref{eq:upper_limit}).

Figure \ref{fig:eff2} shows the variation of polarizing efficiency with extinction. The decreasing trend of polarizing efficiency with extinction, as observed in our study, is similar to that found by many past studies \citep{1992ApJ...389..602J, 1994MNRAS.268....1W, 2013A&A...556A..65E, 2015MNRAS.448.1178H, 2015AJ....149...31J, 2017ApJ...849..157W, 2018MNRAS.475.5535I, 2020AJ....160..256S, 2020AJ....159...99S}. At higher extinction, the less efficiency of polarization indicates that the alignment mechanism for grains is weaker in high extinction regions. A power law decay of polarizing efficiency with extinction is seen in studies \citep{2001ApJ...547..872W, 2008ApJ...674..304W, 2014A&A...569L...1A,2015ARA&A..53..501A, 2015MNRAS.448.1178H, 2015AJ....149...31J, 2019ApJ...873...87M,2021AJ....161..149S, 2022MNRAS.513.4899S}, therefore, we fitted the power law between these parameters and the same is shown in log-log scale in Figure \ref{fig:eff2}. Blue circles show the observed data points, and an orange line in the log-log scale shows the fitted power law. 
The residuals for each data point from the fitted curve are shown at the bottom of the same figure. Only one data point (\#4) has a large residual. The power law relation $P_{\rm max}/A_{V}=(1.46\pm0.08) A_{V}^{(-0.61\pm0.13)}$ was found to be followed in the region of the cluster NGC 7380. The efficiency of alignment in the region can be inferred from the value of the power law exponent, e.g. an exponent of -1 shows negligible alignment efficiency. In contrast, a -0.5 exponent comes due to the magnetic field turbulence, as found in many studies.

\subsection{The Variation of $\lambda_{\rm max}$ and Polarizing Efficiency}
\begin{figure*}
    \centering
    \subfigure[$\lambda_{\rm max}$ with $A_{V}$]{\includegraphics[width=0.48\textwidth]{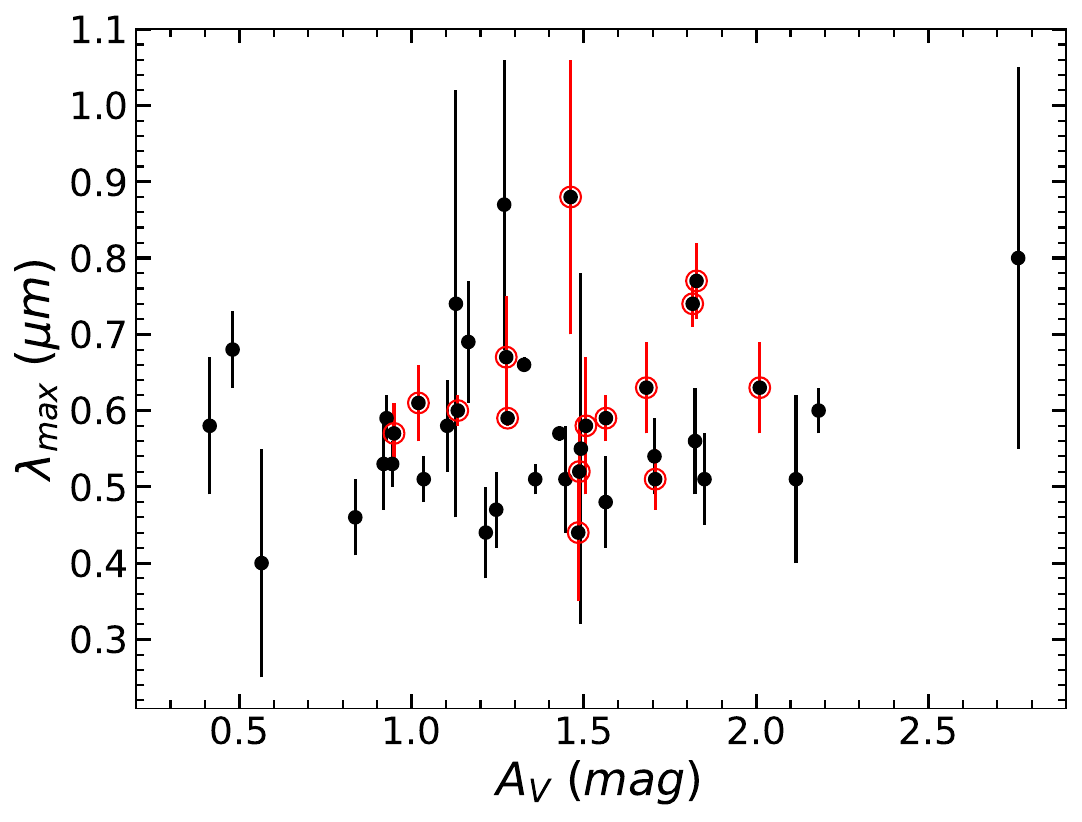} \label{fig:av_lmax}} 
   \subfigure[$\lambda_{\rm max}$ with $P_{\rm max}/A_{V}$]{\includegraphics[width=0.48\textwidth]{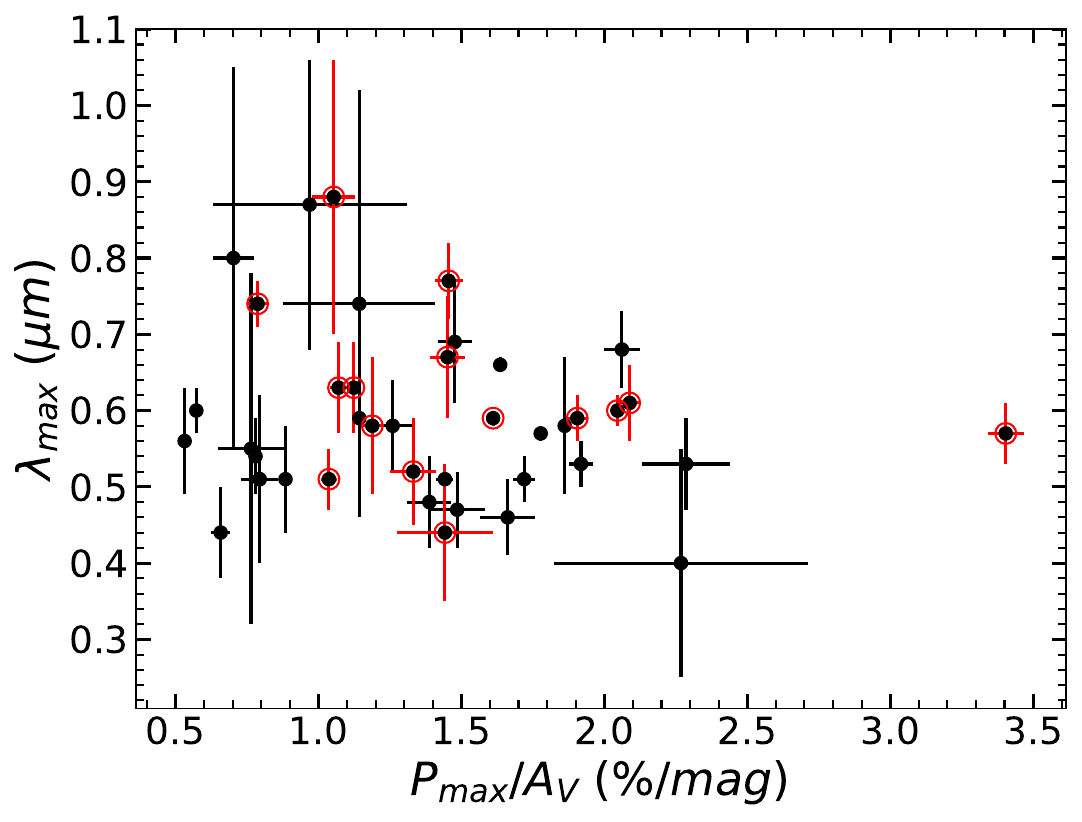} \label{fig:eff_lmax}} 
\subfigure[$P_{\rm max}/A_{V}$ with radial distance]{\includegraphics[width=0.48\textwidth]{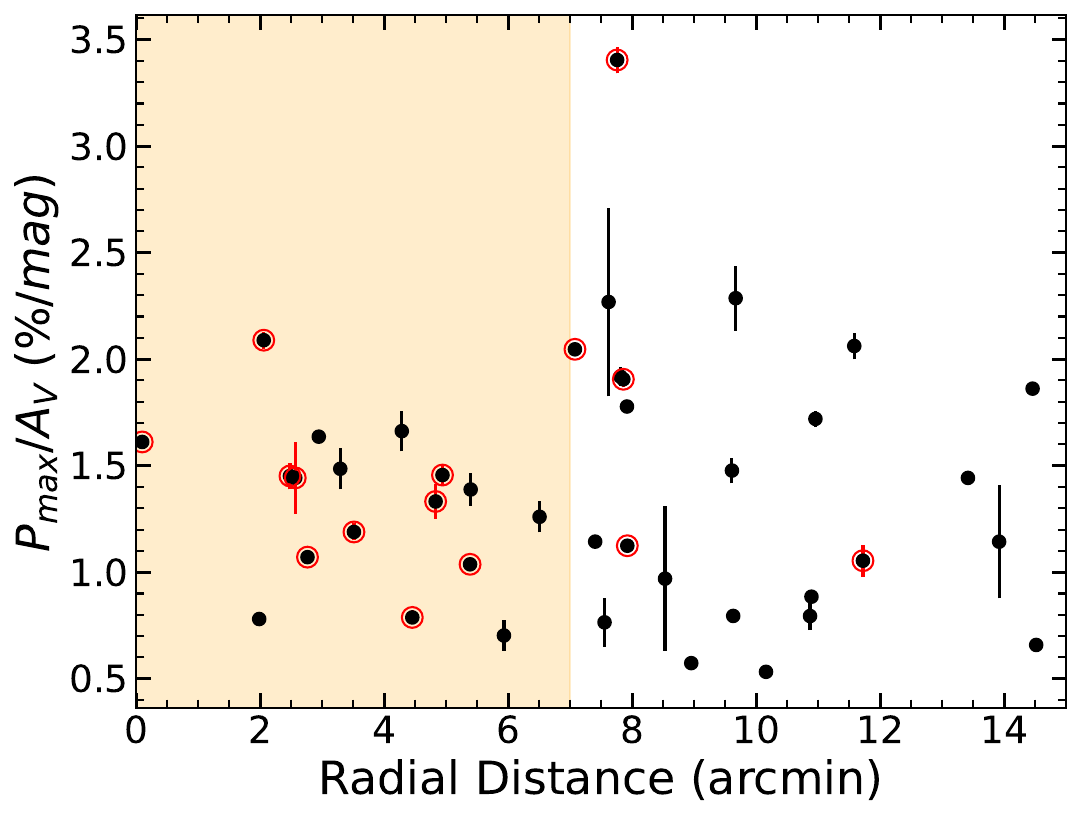} \label{fig:radial_eff}} 
   \subfigure[$\lambda_{\rm max}$ with radial distance]{\includegraphics[width=0.48\textwidth]{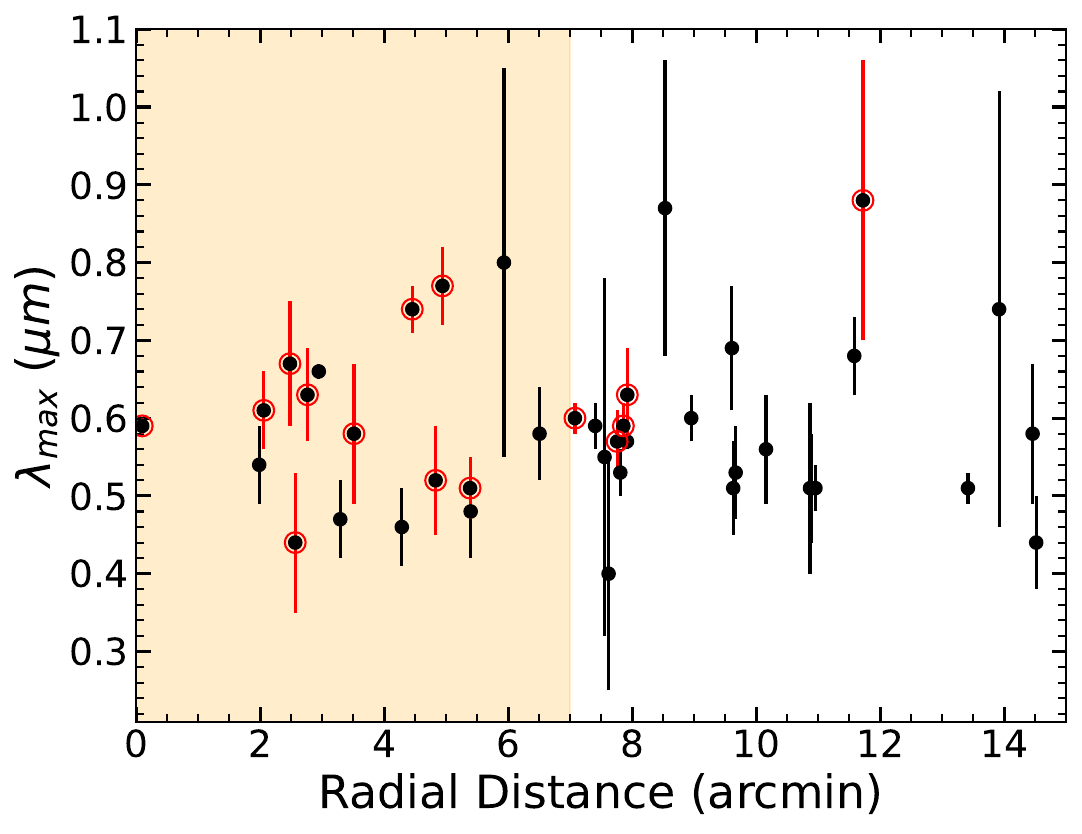}  \label{fig:radial_lmax}}
    \caption{(a), (b): Variation of $\lambda_{\rm max}$ with $A_{V}$ and $P_{\rm max}/A_{V}$. (c), (d): The variation of $P_{\rm max}/A_{V}$ and $\lambda_{\rm max}$ from the centre to outwards of the cluster NGC 7380. The member stars are denoted by the red circles in all panels.}  
    \label{fig:align2}
\end{figure*}

Figure \ref{fig:av_lmax} shows the variation of $\lambda_{\rm max}$ with extinction. It shows a slight increase in $\lambda_{\rm max}$ with increasing $A_{V}$, which can be described by $\lambda_{\rm max} = (0.54\pm0.05) + (0.04\pm0.03) \times A_{V}$. 
There have been studies where $\lambda_{\rm max}$ is found to be weakly correlated with $A_{V}$ \citep[e.g.][]{2001ApJ...547..872W,2007ApJ...665..369A,2008ApJ...674..304W,2018MNRAS.475.5535I, 2022MNRAS.513.4899S}. For our data sets, we found a Pearson correlation coefficient ($r$) between $\lambda_{\rm max}$ and $A_V$ of 0.23 with a probability of no correlation of 0.15, indicating that the relation is not statistically significant.
Numerical modelling for starlight polarization using the RAT theory predicted the increase in the $\lambda_{\rm max}$ with $A_{V}$ toward dense molecular clouds without embedded sources due to the loss of alignment of small grains and grain growth \citep{2008ApJ...674..304W,2015MNRAS.448.1178H,2020ApJ...905..157V,2020ApJ...896...44L}. However, the local environment of star clusters radically differs from dense starless clouds when the main radiation source for RAT alignment is the diffuse interstellar radiation field. This may result in the weak correlation of $\lambda_{max}$ with $A_{V}$ observed in this paper. 

The variation of $\lambda_{\rm max}$ with $P_{\rm max}/A_{V}$ is shown in Figure \ref{fig:eff_lmax}, where the decrease in $\lambda_{\rm max}$ is appeared with the increase in $P_{\rm max}/A_{V}$. As the polarizing efficiency and, hence, the alignment efficiency increases, the average size of aligned grains reduces as smaller grains also align with the increased alignment efficiency. The degree of alignment is predicted to be a function of grain size. In marginal conditions, the smaller grains lose the alignment first, and smaller grains polarize most efficiently at shorter wavelengths; hence, $\lambda_{\rm max}$ shifts towards the longer side in case of lesser alignment efficiency.  
However, a sharp decrement in $\lambda_{\rm max}$ with $P_{\rm max}/A_{V}$ can be expected with observations in a wide range of extinction. In this case, we don't have such a wide range of extinction.  

Usually, factors like massive stars being more concentrated at the cluster centre, their shorter lifespans, radiation pressure, and dynamical processes within the cluster all contribute to the centre of an open star cluster having a higher radiation field than its outer regions. And if the cluster centre is the peak of the radiation field, then the variation of $P_{\rm max}/A_{V}$ with the radial distance from the cluster centre tells us about the alignment efficiency with respect to the local radiation field. Following this approach, we have shown the variation of $P_{\rm max}/A_{V}$ and $\lambda_{\rm max}$ with the radial distance from the cluster centre to outwards in Figure \ref{fig:radial_eff} and \ref{fig:radial_lmax}. 
We found a slight increase of $P_{\rm max}/A_{V}$ and a slight decrease in $\lambda_{\rm max}$ as going towards the centre from 7\arcmin\, distance from the centre of the cluster (the region is highlighted in figures). Beyond a distance of 7\arcmin, there is a noticeable increase in the dispersion of data points. However, the correlation was not found to be strong.
A higher radiation field in the cluster centre causes an increase in polarizing efficiency. As a result, the average size of aligned grains reduces as going toward the cluster centre.

\section{Discussion} \label{diss}
We have carried out the detailed polarization properties of 72 stars towards the direction of cluster NGC 7380, which is associated with emission nebulosity and dark molecular cloud. 
A single peak distribution in polarization and position angle indicates that the source of polarization for stars observed in the direction of the cluster NGC 7380 comes through a similar type of medium, i.e. the interstellar medium. We did not notice any significant turbulence in the plane of the sky component of the magnetic field along the line of sight. 

\subsection{Alignment and Dust Distribution in the Cluster}
Polarization vectors for stars are found to be nearly parallel to  GP, indicating that the magnetic field has played a fundamental role in the alignment of dust grains in the general ISM. Usually, the dust grains are aligned with their short axis parallel to the magnetic field, irrespective of their alignment mechanism. The observed polarization is the integrated polarization through the entire line of sight. The polarization, here, traces the line of sight average of the plane of sky component of the magnetic field.  The magnetic field responsible for the alignment could be either the Galactic field or the local field of the particular region. In the case of NGC 7380, the alignment of dust grains is due to the Galactic magnetic field, as no other effective local fields are present towards the observed region. Members and field stars were found to have the same position angle parallel to the Galactic magnetic field, similar to that found in \citet{1976AJ.....81..970M}. The observed alignment is the magnetic alignment, which is in agreement with the large-scale magnetic field in the Galaxy. 
The H{\sc ii} region (Sh 2-142) is ionized by the principal ionizing star DH Cep (a spectroscopic binary: HD 215835). According to radiative alignment theory, the direction of alignment changes from magnetic to radiation if a strong anisotropic radiation source is nearby \citep{2015ARA&A..53..501A}. Considering this radiation source, we have not found any alignment change. In some past studies, the alignment is found to be different from magnetic alignment near the ionizing source of H{\sc ii} region \citep[e.g.][]{2021AJ....161..149S}. The binary star DH Cep was found to have an intrinsic polarization component of 0.6-1$\%$ with the amplitude of polarization modulation of 0.2-0.7$\%$ and the average polarization angle as 83\degr \citep{2019BSRSL..88..287A, 2023arXiv230517259Q}. The intrinsic nature of polarization in this system arises from the asymmetric circumstellar scattering envelope \citep{1997ApJ...483..439P}. 
 
Dense dust structures in different wavelengths $DSS2-R, WISE-W3$ are visible in the southeast region of the cluster NGC 7380. The location of observed members in the cluster is found to be well correlated with the dense dust structures, which signifies the role of dust in the formation of stars.   
The study led by \citet{1978A&A....70..769I} and \citet{1989ApJS...70..731L} has discussed that the cluster NGC 7380 holds the inherent cloud in the south and east region, which was revealed with the molecular line observations.

\subsection{Size Distribution and Foreground Dust Concentration}
The size distribution of dust grains \citep[as indicated by $\lambda_{\rm max}$;][]{2020ApJ...905..157V} towards the line of sight is similar to that of the general diffuse ISM, i.e. the average value for $\lambda_{\rm max}$ is found to be 0.56$\pm$0.07 $\mu m$, with the variation from 0.40 to 0.88 $\mu m$ in observed stars. In \citet{1976AJ.....81..970M}, the value of $\lambda_{\rm max}$ was not found to be significantly less or more from 0.55 $\mu m$ using ten stars multi-colour polarimetry. 

The information on polarization, extinction, and distances is essential to construct the dust distribution along the line of sight. The increase of polarization and extinction with distance shows the dependency of these parameters on the columnar increment of dust content. Several studies had found the dust layer before the cluster \citep[e.g.][]{2007A&A...462..621V,2008MNRAS.388..105M,2010MNRAS.403.1577M,2011MNRAS.411.1418E,2012MNRAS.419.2587E} based on the sudden jumps in polarization and/or extinction. According to our research published in \citet{2020AJ....160..256S}, clusters close to the Galactic plane have indicated the presence of dust layers. The location of the cluster NGC 7380 near the Galactic plane also makes it probable to have a dust layer in front of this cluster, and the expected result is found. An indication of a dust layer at a distance $\sim$ 1.2 $kpc$ foreground to the cluster NGC 7380 is found.  
Additionally, an increase of polarization and extinction with the distance is seen with the higher dispersion towards the larger distances.

\subsection{Interstellar Environment/Surrounding in the line of sight}
An analysis of the Stokes plane led to the distinction of three different clustering of stars with different polarization parameters. However, in all groups (or overall), few stars are far from the clustering.
The distribution of these groups in the polarization versus distance plot is analyzed. 
Most Group 1 stars are found before the distance 1.2 $kpc$ (where we noticed the dust layer), i.e. foreground stars. However, group 2 and 3 stars are located after the location of the dust layer. A slight regular increase of polarization with distance is seen in each group. In the Stokes plane analysis, we noticed that Group 1 stars are near the solar neighbourhood and have a similar orientation to GP. With the distance, we found that these stars are mostly dispersed foreground to the dust layer with similar $\theta_{V}$ and $P_{V}$ within $\sim$ 1$\pm$0.6$\%$. It shows that the common dust cloud covering the whole line of sight is responsible for the polarization of $\sim$ 1.0$\pm$0.6$\%$ and alignment parallel to GP. Group 2 and 3 stars are located after the dust layer resulting in higher polarization than Group 1. It appears that the dense cloud covering Group 2 stars has very similar orientations to that of Group 1, whereas Group 3 stars are covered with a less dense cloud and have a deviated orientation from GP by $\approx19\degr$. Therefore, after the dust layer, two clouds appear to be present, one of whose orientation is similar to GP and another with a slightly deviated orientation from GP. 
The degree of polarization in Group 3 is generally in the range of 1.4-1.8 \%, whereas Group 2 stars have higher polarization.  The possibility of depolarization for Group 3 stars could be due to the deviation in orientations of the plane of sky magnetic field component along the line of sight.  
If we see the sky positions of these groups of stars, most Group 2 stars are located in the southeast region where the dust structure/cloud is apparently visible. The other two groups' stars are scattered in the plane of the sky.

\subsection{Alignment Efficiency in the Region} 
The polarizing efficiency depends on the ISM dust grains and the Galactic magnetic field properties. 
For the cluster NGC 7380, we estimated the average reddening as $\sim$0.48 mag, which leads to the upper limit for maximum polarization as 4.5$\%$ (Equation \ref{eq:upper_limit}). The average value of $P_{\rm max}$ for the observed region, estimated in Section \ref{sec:ser}, is much less than this upper limit. For the cluster, NGC 7380 to have the average polarizing efficiency ($P_{\rm max}/A_{V}$) similar to the average efficiency for the Galaxy (Equation \ref{eq:av_eff}), the $P_{\rm max}$ should be 1.95$\%$ using the average reddening for the cluster. The distribution of $P_{\rm max}$ of stars is found to have a nearby distribution around this value, as mentioned in Section \ref{sec:ser}, the average value of $P_{\rm max}$ $\approx$ 1.7$\%$. The polarizing efficiency in this region is found to be close to that of the average efficiency for the Galaxy. 
However, the scattering of data points between the maximum efficiency line and lower bound limits (Figure \ref{fig:eff1}) implies the difference in the inclination angle of magnetic fields toward these stars and/or the difference in the grain alignment efficiency \citep{2023arXiv231017048H}.

Variation of polarizing efficiency with extinction is important to verify different models of grain alignment mechanisms. Polarizing efficiency follows a power law decay with extinction with the power law exponent as -0.61 for the region of cluster NGC 7380. A higher value of power-law exponent shows lesser grain alignment efficiency as it goes deeper into the extinct region. For the inner region of the cloud ($A_{V} \geq$ 20 $mag$), the value of power law exponent reaches -1 \citep{2015AJ....149...31J}.

The prevailing theory that accounts for most observed phenomena is the radiative alignment theory. The theory predicts the increase of $\lambda_{\rm max}$ with $A_{V}$ due to the attenuation of the interstellar radiation field toward denser regions (higher $A_{V}$). In the regime of strong radiation sources, the increase in the grain alignment with the radiation strength accompanied by a shift in the peak wavelength ($\lambda_{\rm max}$) towards shorter values is predicted by RAT theory \citep{2020ApJ...896...44L}.  In this study, the peak wavelength seems to have a weak correlation with $A_{V}$ (see Figure \ref{fig:av_lmax}), which may arise from the fact that the local properties such as the gas density and radiation field toward this special region of open star clusters are not significantly varying with $A_{V}$, unlike in dense molecular clouds.
We also found observational evidence for the dependence of grain alignment on factors like the size of grains (Figure \ref{fig:eff_lmax}), the radiation ﬁeld (Figure \ref{fig:radial_eff}, \ref{fig:radial_lmax}), the density of regions or extinction (Figure \ref{fig:eff2}), which are consistent with the predictions of RAT theory \citep[][etc.]{2005ApJ...631..361C, 2007MNRAS.378..910L, 2008MNRAS.388..117H, 2015MNRAS.448.1178H, 2015ARA&A..53..501A}.

\section{Conclusions and summary} \label{summ}
\label{summ}
We study the properties of foreground dust toward the star cluster NGC 7380 using the multi-band polarimetric data observed with a 104-cm ARIES telescope. Our main results are summarized as follows:
\begin{itemize}
    \item 
A single peak distribution of polarization and position angle is found in all four bands, specifying the same source for polarization responsible for members and non-members of the cluster, and is the foreground dust. A higher polarization towards the east and southeast regions of the cluster is noticed, indicating a higher dust concentration in that particular region, as inferred from the extinction contours and colour composite image of the region. 
 \item
The majority of polarization vectors are found to be parallel to the GP, which implies that the alignment of dust grains in the region is with the Galactic magnetic field, i.e. magnetic alignment. 
 \item
By analyzing the variation of the polarization degree and extinction with distance, we found the presence of a dust layer at a distance of $\sim$1.2 kpc. However, there is no particular change in the position angle at this distance. 
\item
Using three criteria ($\sigma_{1}$, $\lambda_{\rm max}$, and $\bar{\epsilon}$) for the differentiation between intrinsic and interstellar polarized stars, eight probable candidates are found for intrinsic polarization. This selection of stars represents promising cases for further investigation into their inherent polarization properties.
 \item
The maximum interstellar polarization and corresponding wavelength are 1.71$\pm$0.57$\%$ and 0.56$\pm$0.07 $\mu m$, indicating the similar size distribution of dust grains as in the diffuse ISM. 
\item
The polarizing efficiency of dust grains is found to be close to the average polarizing efficiency for the Galaxy with the dispersion of observed stars from lower to upper bound on polarizing efficiency. Polarizing efficiency followed a power law decay with extinction with the power law exponent as -0.61$\pm$0.13. The variation of polarizing efficiency with the extinction, radiation strength, and size of grains are in agreement with the prediction of radiative torque alignment theory. 
\end{itemize}

\begin{acknowledgments}
\section*{acknowledgments}
We thank the referee for providing useful comments and suggestions that helped to improve the paper. Part of this work has made use of data from the European Space Agency (ESA) mission Gaia (https://www.cosmos.esa.int/gaia), processed by the Gaia Data Processing and Analysis Consortium (DPAC, https://www.cosmos.esa.int/web/gaia/dpac/consortium). Funding for the DPAC has been provided by national institutions, in particular, the institutions participating in the Gaia Multilateral Agreement. Work at the Physical Research Laboratory is supported by the Department of Space, Govt. of India. TH acknowledges the support of the National Research Foundation of Korea (NRF) grants funded by the Korean government (MSIT) through the Mid-career Research Program (2019R1A2C1087045). SS acknowledges Priyanka Srivastava for help during observations. NP acknowledges the financial support received through the SERB CRG/2021/005876 grant. 
\end{acknowledgments}


\bibliography{ref2}{}
\bibliographystyle{aasjournal}

\end{document}